\documentclass[aps, preprint, showpacs, amsmath, amssymb, pra, 12pt]{revtex4}
\usepackage{graphicx}% Include figure files
\usepackage{dcolumn}% Align table columns on decimal point
\usepackage{color}
\usepackage{bm}% bold math
\begin{document}
%\begin{CJK*}{UTF8}{bsmi}
\title{Fully quantum-mechanical analytic results for single-photon transport in a single-mode waveguide coupled to a whispering-gallery resonator interacting with a two-level atom}
\author{Jung-Tsung Shen}
\email{jushen@stanford.edu}
\author{Shanhui Fan}
\email{shanhui@stanford.edu}
\affiliation{Ginzton Laboratory, Stanford University, Stanford, CA
94305}
\date{\today}
\begin{abstract}
We analyze the single-photon transport in a single-mode waveguide coupled to a whispering-gallery-type resonator interacting with a two-level atom. The single-photon transport properties such as the transmission and reflection amplitudes, as well as the resonator and the atom responses, are solved exactly via a real-space approach. The treatment includes the inter-mode backscattering between the two degenerate whispering gallery modes of the resonator, and the dissipations of the resonator and the atom. We also show that a generalized critical coupling condition, that the single-photon transmission at the output of the waveguide goes to zero on resonance for a matched system, holds for the full coupled waveguide-ring resonator-atom system.
\end{abstract}
 \maketitle
%\end{CJK*}

\section{Introduction}

\begin{figure}[thb]
\scalebox{1}{\includegraphics{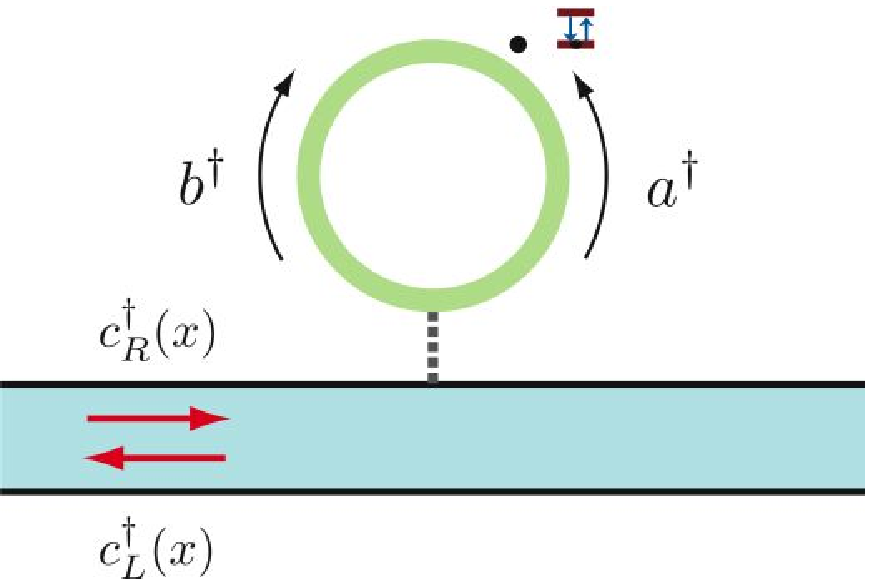}}
%\scalebox{1}{\includegraphics[width=\columnwidth]{Cavity_Waveguide_3.eps}}
\caption{(Color online) Schematics of the system. The single-mode waveguide is denoted by the blue channel. The ring resonator is denoted by the green ring. The black dot denotes the two-level atom, of which the two energy levels are plotted on the right.  The waveguiding modes are described by $c_R^{\dagger}(x)$ and $c_L^{\dagger}(x)$. The two degenerate whispering gallery modes are described by $a^{\dagger}$(counter-clockwise) and $b^{\dagger}$(clockwise).}\label{Fi:Schematics}
\end{figure}

There has been a lot of recent interest in the systems of coupled whispering-gallery-type micro-resonators and atoms~\cite{Thompson:1992a,Vernooy:1998,Klimov:1999, Rosenblit:2004, Aoki:2006,  Srinivasan:2007, Srinivasan:2007a,Mazzei:2007, Dayan:2008} (a schematic configuration is shown in Fig.~\ref{Fi:Schematics}). This configuration is particularly explored in a variety of fundamental and applied studies such as cavity quantum electrodynamics (cavity QED), single-atom manipulation and detection, and biochemical sensing. In the context of cavity QED, this kind of system has been treated semiclassically by assuming a weak classical input~\cite{Carmichael:2003}. Recently a comprehensive study of this system based on the semiclassical approach has been carried out by Srinivasan and Painter~\cite{Srinivasan:2007}. Such semiclassical treatment is of direct current experimental interest, since the incident state in most cavity QED experiments are indeed weak classical beam. 

There are, however, several reasons for which the semiclassical treatment is not completely satisfactory. First of all, given that these systems would ultimately be used to process quantum states of photons, it would seem valuable to treat these systems with fully quantized input states as well. Secondly, even in the presence of a classical input beam, the output of this system is not a coherent state. The semiclassical treatment typically provides various correlation functions of the output state. For a complete description of the output, however, one might instead prefer a direct description of the output quantum state itself. Thirdly, the semiclassical treatment is typically carried out numerically, by setting up a set of master equations for the evolution of the density matrix for the atom-cavity system, and by truncating the density matrix assuming a maximum number of photons in the cavity. For very weak classical input, for example, the maximum number of photon is often taken to be 1. For a deeper understanding of the underlying physics, however, exact analytic results are quite valuable. Finally, the physics intuition related to a semiclassical treatment has to be always treated with caution. For example, the truncation procedure, as discussed above, yields a set of coupled nonlinear ordinary differential equations, even when the maximum photon number in the cavity is fixed to be 1. Based upon this set of coupled nonlinear differential equations, one may be tempted to discuss the so-called ``single-photon nonlinearity''. However, it is worthwhile to emphasize that  nonlinearity has to arise from photon-photon interaction.

Motivated by the discussions above, in this article, we consider the single-photon transport for the system of coupled whispering-gallery-type micro-resonators and atoms, assuming a single-photon Fock state input. In contrast to the semiclassical treatments, where in most cases only numerical results are obtained so far, here the use of a fully quantized formalism allows one to straightforwardly obtain  analytic and exact results of the transport properties and the system responses, which has not been done before for this system. Our analytic results are in well agreement with the experimental data~\cite{Kippenberg:2004, Aoki:2006} and with the numerical results in Srinivasan and Painter's for the parameters considered in their studies~\cite{Srinivasan:2007}. Using these analytic results, we systematically explore the parameter regime that is of direct experimental interests, to understand the delicate interplay between the various types of coupling and the intrinsic atom and resonator dissipations.

This article is organized as follows. In Sec.~\ref{Sec:System} we first introduce the system configuration and the Hamiltonian. In Sec.~\ref{Sec:Solutions}, the single-photon transport properties and the system responses are then solved analytically. Sec.~\ref{Sec:GeneralProperties} describes the general properties of the exact solutions, such as spectral symmetry of the transmission spectrum, and generalized critical coupling condition of the full coupled system. In Sec.~\ref{Sec:Numerical}, we compare our results with some experimental results of waveguide-resonator systems, and with published numerical results of coupled waveguide-resonator-quantum dot systems.
% In Sec.~\ref{Sec:WR}, we discuss the case of coupled waveguide-ring resonator without the atom. The effects of cavity dissipation and mode cross-talking are investigated. 
Sec.~\ref{Sec:WRA} presents a systematic study of the single-photon transport of the full coupled system of waveguide-ring resonator-atom. Sec.~\ref{Sec:Dissipations} discusses the effects of the intrinsic dissipations of the resonator and the atom. In Sec.~\ref{Sec:Complexh} we investigate the effects of complex inter-mode backscattering. Sec.~\ref{Sec:Detuning} discusses the effects of detuning between the resonator and the atom. %Sec.~\ref{Sec:GeneralizedCritical} describes the generalized critical coupling condition with a concrete example. 
Finally, Sec.~\ref{Sec:Summary} sums up the article, pointing out some applications and generalizations of the formalism.

\section{The system and Hamiltonian}\label{Sec:System}

%\begin{figure}[thb]
%\scalebox{1}{\includegraphics{Geometry.eps}}
%%\scalebox{1}{\includegraphics[width=\columnwidth]{Cavity_Waveguide_3.eps}}
%\caption{(Color online) Schematics of the system. The single-mode waveguide is denoted by the blue channel. The ring resonator is denoted by the green ring. The black dot denotes the two-level atom. The waveguiding modes are described by $c_R^{\dagger}(x)$ and $c_R^{\dagger}(x)$. The two degenerate whispering gallery modes are described by $a^{\dagger}$(counter-clockwise) and $b^{\dagger}$(clockwise).}\label{Fi:Schematics}
%\end{figure}

The system of interest in this article is schematically shown in Fig.~(\ref{Fi:Schematics}): a whispering-gallery type resonator interacting with a two-level atom is side-coupled to a single-mode waveguide. A whispering-gallery type microresonator, such as a ring resonator, a microsphere, a microtoroid or microdisk, supports two degenerate whispering gallery modes (WGMs) that propagate around the resonator in opposite directions. (Throughout this paper, we will often use the term ``ring resonator'' to designate a whispering-gallery type microresonator). We also include the interactions of the atom and the resonator with the reservoirs. Such interactions with reservoirs give rise to intrinsic dissipations~\cite{Scully:1997, Carmichael:2003}.
%The Hamiltonian of the system is described as
%a ring resonator interacting with a two-level atom is side-coupled to a single-mode waveguide. 
%The ring resonator supports a pair of degenerate, counter-propagating whispering gallery modes in the frequency of interest. The Hamiltonian of the system is described as
%%%%%% V %%%%%%%
%\begin{align}\label{E:Hamiltonian}
%H/\hbar &=\int dx\, c^{\dagger}_R(x)\left(\omega_0-i v_g \frac{\partial}{\partial x}\right)c_R(x) + \int dx \,c^{\dagger}_L(x)\left(\omega_0+i v_g \frac{\partial}{\partial x}\right)c_L(x)\notag\\
%&+ \left(\omega_c -i\frac{1}{\tau_c}\right) a^{\dagger}a + \left(\omega_c -i\frac{1}{\tau_c}\right) b^{\dagger}b+\left(\Omega_e - i\frac{1}{\tau_q}\right) a_e^{\dagger}a_e +  \Omega_g a_g^{\dagger}a_g\notag\\
%&+\int dx \, V^{*}\delta(x)[a^{\dagger}c_R(x) + b^{\dagger}c_L(x)]+ \int dx \, V\delta(x)[c_R^{\dagger}(x)a + c_L^{\dagger}(x) b]\notag\\
%&+\left(g_a a\sigma_{+} + g_a^* a^{\dagger}\sigma_{-}\right) + \left(g_b b\sigma_{+} + g_b^* b^{\dagger}\sigma_{-}\right)\notag\\
%%&+\left(g \sigma^{+}a + g^{*}a^{\dagger}\sigma^{-}\right) + \left(g^{*} \sigma^{+}b+ g b^{\dagger}\sigma^{-}\right)\notag\\
%&+ \left(h b^{\dagger}a +h^{*}a^{\dagger}b\right).
%\end{align}
%%%%% V_a V_b %%%%%%

The Hamiltonian of the \emph{composite} system $S\bigoplus R$ is $H\equiv H_S + H_R + H_{SR}$:
\begin{subequations}
\begin{align}
H_{S}/\hbar &=\int dx\, c^{\dagger}_R(x)\left(\omega_0-i v_g \frac{\partial}{\partial x}\right)c_R(x) + \int dx \,c^{\dagger}_L(x)\left(\omega_0+i v_g \frac{\partial}{\partial x}\right)c_L(x)\notag\\
&+ \omega_c  a^{\dagger}a + \omega_c  b^{\dagger}b+\Omega_e  a_e^{\dagger}a_e +  \Omega_g a_g^{\dagger}a_g\notag\\
&+\int dx \, \delta(x)[V_a c_R^{\dagger}(x)a + V_a^* a^{\dagger} c_R(x)] + \int dx \, \delta(x)[V_b c_L^{\dagger}(x)b + V_b^* b^{\dagger} c_L(x)]\notag\\
&+\left(g_a a\sigma_{+} + g_a^* a^{\dagger}\sigma_{-}\right) + \left(g_b b\sigma_{+} + g_b^* b^{\dagger}\sigma_{-}\right)\notag\\
&+ \left(h b^{\dagger}a +h^{*}a^{\dagger}b\right),\label{E:HS}\\
H_{R}/\hbar &= \sum_j \omega_{1j} r^{\dagger}_{1j} r_{1j} + \sum_j \omega_{2j} r^{\dagger}_{2j} r_{2j} + \sum_j \omega_{3j} r^{\dagger}_{3j} r_{3j},\label{E:HR}\\
H_{SR}/\hbar &= \sum_j \left(\kappa^{*}_{j} r^{\dagger}_{1j} a + \kappa_{j} a^{\dagger} r_{1j}\right)  + \sum_j \left(\kappa^{*}_{j} r^{\dagger}_{2j} b + \kappa_{j} b^{\dagger} r_{2j}\right) + \sum_j \left(\eta^{*}_{j} r^{\dagger}_{3j} \sigma_{-} + \eta_{j} \sigma_{+} r_{3j}\right).\label{E:HSR}
\end{align}
\end{subequations}

$H_S$ is the Hamiltonian of the system $S$ of coupled waveguide-resonator-atom. This Hamiltonian includes the waveguide, the resonator, and the atomic part, as well as the interaction between the waveguide and the resonator, and the atom and the resonator. 

The first line in $H_S$ of Eq.~\eqref{E:HS} describes the propagating photon modes in the waveguide. $c^{\dagger}_{R/L}(x)$ is a bosonic operator creating a right/left-moving photon at $x$. $\omega_0$ is a reference frequency, around which the waveguide dispersion relation is linearized~\cite{Linearization}. Its value does not affect the photon transport properties.  

The second line in $H_S$ describes the modes in the resonators and the atomic states. For the resonator,  $a^{\dagger}$ is the creation operator for the counter-clockwise WGM mode and $b^{\dagger}$ is the creation operator for the clockwise WGM mode,  both of frequency $\omega_c$. For the atom, $a^{\dagger}_{g}$($a^{\dagger}_{e}$) is the creation operator of the
ground (excited) state, $\sigma_{+}=a^{\dagger}_{e}
a_{g}$($\sigma_{-}=a^{\dagger}_{g} a_{e}$) is the atomic raising (lowering) ladder
operator satisfying $\sigma_{+}|n,n_c=0,-\rangle =|n,n_c=0, +\rangle$ and $\sigma_{+}|n,n_c,+\rangle =0$, where $|n, n_c, \pm\rangle \equiv |n\rangle\otimes|n_c\rangle\otimes|\pm\rangle$ describes the state of the system with $n$ propagating photons, $n_c$ photons in cavity mode, and the atom in the excited ($+$) or ground ($-$) state. $\Omega_{e}-\Omega_{g}(\equiv\Omega)$ is the atomic transition frequency. 

The third line in $H_S$ describes the interactions between the waveguiding modes and the WGMs. $V_{a/b}$ is the waveguide-resonance coupling strength of each WGM. Here the right-moving(left-moving) mode only couples to the phase-matched counter-clockwise(clockwise) mode. 

The fourth line in $H_S$ describes the interactions between the atom and the whispering gallery modes. $g_a$ and $g_b$ are the resonator-atom coupling strength for each respective WGM.  

The last line in $H_S$ describes the inter-mode backscattering between the two degenerate WGMs, induced through imperfection of the resonator. $h$ is the inter-mode backscattering strength.  

$H_R$ of Eq.~\eqref{E:HR} describes the reservoir, which is composed of three subsystems: $R=R_1 \bigoplus R_2 \bigoplus R_3$. $R_1$, $R_2$ and $R_3$ are assumed to be independent. Each $R_{1}$, $R_2$ and $R_3$ is modeled as a collection of harmonic oscillators with frequencies $\omega_{1j}$, $\omega_{2j}$, and $\omega_{3j}$, and with the corresponding creation (annihilation) operators $r_{1j}^{\dagger}$ ($r_{1j}$), $r_{2j}^{\dagger}$ ($r_{2j}$), and $r_{3j}^{\dagger}$ ($r_{3j}$), respectively.

$H_{SR}$ of Eq.~\eqref{E:HSR} describes the interactions between the resonator and the atom with the reservoirs, respectively. The WGM $a^{\dagger}$ couples to the $j$th reservoir oscillator $r_{1j}$ in $R_1$, while the WGM $b^{\dagger}$ couples to the $j$th reservoir oscillator $r_{2j}$ in $R_2$. Both WGMs are assumed to couple with the reservoir with the same coupling constant $\kappa_{j}$. The atom $\sigma_{+}$ couples to the $j$th reservoir oscillator $r_{3j}$ in $R_3$ with a coupling constant $\eta_j$. 

By incorporating the excitation amplitudes of the reservoir $R$, it can be shown that the effective Hamiltonian $H_{\text{eff}}$ of $S$ can be obtained and is given by~\cite{Linearization}:
\begin{align}\label{E:Hamiltonian}
H_{\text{eff}}/\hbar &=\int dx\, c^{\dagger}_R(x)\left(\omega_0-i v_g \frac{\partial}{\partial x}\right)c_R(x) + \int dx \,c^{\dagger}_L(x)\left(\omega_0+i v_g \frac{\partial}{\partial x}\right)c_L(x)\notag\\
&+ \left(\omega_c -i\frac{1}{\tau_c}\right) a^{\dagger}a + \left(\omega_c -i\frac{1}{\tau_c}\right) b^{\dagger}b+\left(\Omega_e - i\frac{1}{\tau_q}\right) a_e^{\dagger}a_e +  \Omega_g a_g^{\dagger}a_g\notag\\
&+\int dx \, \delta(x)[V_a c_R^{\dagger}(x)a + V_a^* a^{\dagger} c_R(x)] + \int dx \, \delta(x)[V_b c_L^{\dagger}(x)b + V_b^* b^{\dagger} c_L(x)]\notag\\
&+\left(g_a a\sigma_{+} + g_a^* a^{\dagger}\sigma_{-}\right) + \left(g_b b\sigma_{+} + g_b^* b^{\dagger}\sigma_{-}\right)\notag\\
%&+\left(g \sigma^{+}a + g^{*}a^{\dagger}\sigma^{-}\right) + \left(g^{*} \sigma^{+}b+ g b^{\dagger}\sigma^{-}\right)\notag\\
&+ \left(h b^{\dagger}a +h^{*}a^{\dagger}b\right),
\end{align} where $1/\tau_c \equiv \gamma_c$ and  $1/\tau_a\equiv \gamma_a$ are the intrinsic dissipation rates of the resonator WGM and the atom, respectively, due to coupling to the reservoir. We will call $H_{\text{eff}}$ as $H$ in the following for brevity. Note that $|V_{a/b}|^2/v_g$, $1/\tau_c$, $1/\tau_q$, $g_a$, $g_b$, and $h$ all have the same unit as frequency.

Here we make some remarks on the Hamiltonian of Eq.~\eqref{E:Hamiltonian}:
\begin{enumerate}
\item The coupling constants are determined by the underlying photonic and electronic states of each constituent of the system. The relations between these coupling constants are not arbitrary but rather are constrained by symmetry of the system. These symmetry properties in turn are reflected in the response spectra of the system. The relevant symmetry operations here are mirror and time-reversal symmetries. A discussion of the symmetry transformations is given in Appendix~\ref{A:SymmetryTransformations}. 

 %In particular, one always has $|V_a|=|V_b|$. A discussion of the symmetry constraints is given in Appendix~\ref{A:RAInteraction}. 
\item Except obeying the symmetry transformations, the specific numerical values of the coupling constants depend upon each specific experimental realization. A detailed discussion is provided in Appendix~\ref{A:RAInteraction}. Here, we just note that that in general $|g_a| \neq |g_b|$, and both $g_a$ and $g_b$ can be complex numbers.

%For example, in the study of Srinivasan and Painter's on the system of waveguide-microdisk-quantum dot~\cite{Srinivasan:2007}, one could have $g_b = g_a$ ($\equiv -i g_0$ in their paper) by choosing the azimuthal origin lying at the location of the quantum dot, and a quantum dot dipole polarization transverse to the azimuthal direction. On the other hand, in Refs.~[\onlinecite{Aoki:2006,Dayan:2008}] where the transitions are between the hyperfine states of an atom, one could have $g_a = g_b^*$  ($\equiv g_{\text{TW}}$ in their papers) if both WGMs are transverse electric (TE) mode so the dipole matrix element is purely real. In both cases, one has $|g_a| = |g_b|$.  In general, however, $|g_a|$ can be different from $|g_b|$ in systems that do not have time-reversal symmetry (see Appendix~\ref{A:RAInteraction}).
\end{enumerate}

In the following, we first derive the exact solutions to the single-photon transport and the system responses (\emph{i.e.}, WGMs and atom excitations) described by the Hamiltonian of Eq.~\eqref{E:Hamiltonian}, without placing any constraints on the coupling constants. Later, we will specialize to specific choices of the coupling constants to compare with experimental data, and to some published numerical results.

%Here we comment on the reality of the coupling constant $g$ and $h$. For systems consisting of one resonator and one atom, the azimuthal origin could be chosen to be at the location of the atom such that $g$ is purely real. This choice however is not possible for general situations, in particular when several ring resonators are cascaded, each one interacting with one atom each at different angular positions. $h$ is also in general complex. The transport properties, such as the transmission and reflection amplitudes, depend upon the phases of $g$ and $h$.

\section{Exact Solutions for Single-Photon Transport}\label{Sec:Solutions}
We consider the temporal evolution of an arbitrary single-photon state $|\Phi(t)\rangle$, as described by the Schr\"odinger equation
\begin{equation}\label{E:Schrodinger}
i \hbar \frac{\partial}{\partial t}|\Phi(t)\rangle=H|\Phi(t)\rangle,
\end{equation} where $H$ is the Hamiltonian of Eq.~(\ref{E:Hamiltonian}). In general, $|\Phi(t)\rangle$ can be expressed as
\begin{align}\label{E:State}
|\Phi(t)\rangle = & \int dx \left[\tilde{\phi}_R (x, t) c_R^{\dagger}(x) + \tilde{\phi}_L (x, t) c_L^{\dagger}(x)\right]|\emptyset\rangle\notag\\
&+ \tilde{e}_a(t) a^{\dagger}|\emptyset\rangle + \tilde{e}_b(t) b^{\dagger}|\emptyset\rangle + \tilde{e}_q(t) \sigma^{\dagger}|\emptyset\rangle,
\end{align}where $|\emptyset\rangle$ is the vacuum, which has zero photon and has the atom in the ground state. $\tilde{\phi}_{R/L}(x, t)$ is the single-photon wave function in the $R/L$ mode.
$\tilde{e}_{a/b}(t)$ is the excitation amplitude of the whispering gallery mode, and $\tilde{e}_q(t)$ is the excitation amplitude of the atom. For this state, the Schr\"odinger equation (Eq.~(\ref{E:Schrodinger})) thus gives the following set of equations of motion:
\begin{subequations}\label{E:TEoM}
\begin{align}
-i v_g\frac{\partial}{\partial x}\tilde{\phi}_R(x, t) &+ \delta(x) V_a \tilde{e}_a(t) +  \left(\omega_0+\Omega_g\right) \tilde{\phi}_R(x, t)= i\frac{\partial}{\partial t} \tilde{\phi}_R(x, t),\label{E:TEoM1}\\
+iv_g\frac{\partial}{\partial x}\tilde{\phi}_L(x, t) &+ \delta(x) V_b \tilde{e}_b(t) +  \left(\omega_0+\Omega_g\right) \tilde{\phi}_L(x, t)= i\frac{\partial}{\partial t} \tilde{\phi}_L(x, t),\label{E:TEoM2}\\
(\omega_c +\Omega_g-i\frac{1}{\tau_c}) \tilde{e}_a(t) &+ V_a^{*} \tilde{\phi}_R(0, t) + g_a^* \tilde{e}_q(t) + h^* \tilde{e}_b(t) =i\frac{\partial}{\partial t} \tilde{e}_a(t),\label{E:TEoM3}\\
(\omega_c +\Omega_g-i\frac{1}{\tau_c}) \tilde{e}_b(t) &+ V_b^{*} \tilde{\phi}_L(0, t) + g_b^{*} \tilde{e}_q(t)+ h \tilde{e}_a(t) =i\frac{\partial}{\partial t} \tilde{e}_b(t),\label{E:TEoM4}\\
(\Omega_e -i\frac{1}{\tau_q}) \tilde{e}_q(t) &+ g_a \tilde{e}_a(t) + g_b \tilde{e}_b(t) = i\frac{\partial}{\partial t} \tilde{e}_q(t).\label{E:TEoM5}
\end{align}
\end{subequations}For any given initial state $|\Phi(t=0)\rangle$, the dynamics of the system can be obtained directly by  integrating this set of equations (Eqs.~(\ref{E:TEoM})). In this way,  one could study the time-dependent transport of an arbitrary single-photon wave packet.

In the following, we concentrate on the steady state properties.  When $|\Phi(t)\rangle$ is an eigenstate of frequency $\epsilon$, \emph{i.e.}, $|\Phi(t)\rangle = e^{- i \epsilon t}|\epsilon^{+}\rangle$, Eq.~(\ref{E:Schrodinger}) yields the time-independent eigen equation
\begin{equation}\label{E:TimeIndependent}
H|\epsilon^+\rangle =\hbar\epsilon|\epsilon^+\rangle.
\end{equation} and the interacting steady-state solution $|\epsilon^+\rangle$ can be solved for. Here $\hbar\epsilon$ is the total energy of the system.

For an input state of one-photon Fock state, the most general time-independent interacting eigenstate for the Hamiltonian of Eq.~(\ref{E:Hamiltonian}) is:
\begin{equation}\label{E:InteractingEigenstate}
|\epsilon^+\rangle = \int dx\,\left[\phi_R(x) c_R^{\dagger}(x) + \phi_L(x) c_L^{\dagger}(x)\right]|\emptyset\rangle + e_a a^{\dagger}|\emptyset\rangle  + e_b b^{\dagger}|\emptyset\rangle + e_q \sigma^{+}|\emptyset\rangle,
\end{equation}where we denote the time-independent amplitudes by the corresponding untilded symbols, \emph{e.g.} $\tilde{e}_a(t) = e_a \, e^{- i \epsilon t}$, etc. The connection between the interacting eigenstate and a scattering experiment is described by the Lippmann-Schwinger formalism~\cite{Taylor:1972,Huang:1998,Shen:2007d}. 
%For the purpose of this article, one is interested the following scenario: in the remote past ($t\rightarrow -\infty$),  a one-photon state
%\begin{equation}\label{E:FreeState}
%|\epsilon\rangle \equiv \int dx\, e^{i k_R x} c_R^{\dagger}(x)|\emptyset\rangle,
%\end{equation}with total energy $\hbar\epsilon\equiv\hbar\left(\omega+\Omega_g\right)$ is prepared. $\omega =\omega_0 + v_g k_R$ is the frequency of the photon, and $\Omega_g$ is the frequency of the atomic ground state. As proved in Ref.~[\onlinecite{Shen:2007d}], the outgoing one-photon state at the remote future ($t\rightarrow +\infty$) after scattering is given by the $x>0$ part of the interacting eigenstate. Thus, in the following, we focus on solving the eigenstate with the same total energy $\hbar\epsilon\equiv\hbar\left(\omega+\Omega_g\right)$. 

The time-independent Schr\"odinger equation of Eq.~(\ref{E:TimeIndependent}) for the state $|\epsilon^+\rangle$ of Eq.~(\ref{E:InteractingEigenstate}) yields the following equations of motion:
\begin{subequations}\label{E:EoM}
\begin{align}
-i v_g\frac{\partial}{\partial x}\phi_R(x) &+ \delta(x) V_a e_a = \left(\epsilon-\omega_0-\Omega_g\right) \phi_R(x),\label{E:EoM1}\\
+iv_g\frac{\partial}{\partial x}\phi_L(x) &+ \delta(x) V_b e_b = \left(\epsilon-\omega_0-\Omega_g\right) \phi_L(x),\label{E:EoM2}\\
(\omega_c -i\frac{1}{\tau_c}) e_a &+ V_a^{*} \phi_R(0) + g_a^* e_q + h^* e_b =\left(\epsilon-\Omega_g\right) e_a,\label{E:EoM3}\\
(\omega_c -i\frac{1}{\tau_c}) e_b &+ V_b^{*} \phi_L(0) + g_b^{*} e_q + h e_a =\left(\epsilon-\Omega_g\right) e_b,\label{E:EoM4}\\
(\Omega -i\frac{1}{\tau_q}) e_q &+ g_a e_a + g_b e_b = \left(\epsilon-\Omega_g\right) e_q,\label{E:EoM5}
\end{align}
\end{subequations}with $\epsilon=\omega+\Omega_g$, and $\omega=\omega_0 + v_g k_R$.
Our aim is to solve for the transmission and reflection amplitudes for an incident photon. For this purpose, we take $\phi_R(x) = e^{i {Q} x}\left(\theta(-x) + t \theta(x)\right)$, and $\phi_L(x) = r e^{-i {Q} x}\theta(-x)$, where $t$ is the transmission amplitude, and $r$ is the reflection amplitude~\cite{Shen:2005, Shen:2005a}. The set of equations of motion, Eqs.~(\ref{E:EoM1})-(\ref{E:EoM5}) now read:
\begin{subequations}\label{E:EoMTwo}
\begin{align}
-i v_g (t-1) &+  V_a e_a = 0,\label{E:EoM21}\\
+iv_g (-r) &+ V_b e_b =0,\label{E:EoM22}\\
(\omega_c -i\frac{1}{\tau_c}) e_a &+ V_a^{*} \frac{1+t}{2} + g_a^* e_q + h^* e_b =\omega e_a,\label{E:EoM23}\\
(\omega_c -i\frac{1}{\tau_c}) e_b &+ V_b^{*} \frac{r}{2} + g_b^* e_q + h e_a =\omega e_b,\label{E:EoM24}\\
(\Omega -i\frac{1}{\tau_q}) e_q &+ g_a e_a + g_b e_b = \omega e_q,\label{E:EoM25}
\end{align}
\end{subequations}
which can be solved straightforwardly for $Q$, $t$, $r$, $e_c$, and $e_a$:
{\scriptsize
\begin{subequations}\label{E:Amplitudes}
\begin{align}
Q &= \frac{\omega-\omega_0}{v_g},\label{E:Q}\\
t & = \frac{\left(\omega-\omega_c+ i \frac{1}{\tau_c} \right)\left[\left(\omega-\Omega+ i\frac{1}{\tau_q}\right)\left(\omega-\omega_c+ i \frac{1}{\tau_c}\right) -G_{+}^2\right]+\left(\omega-\Omega+ i\frac{1}{\tau_q}\right)\Gamma^2 - g_a^* g_b h -g_a g_b^* h^*-{|h|}^2 \left(\omega -\Omega + i\frac{1}{\tau_q}\right) + iG_{-}^2\Gamma}{ \left(\omega -{\omega_c}+ i \frac{1}{\tau_c}+i \Gamma\right)  \left[\left(\omega-\Omega+ i\frac{1}{\tau_q}\right) \left(\omega -{\omega_c}+ i \frac{1}{\tau_c}+i \Gamma\right)-G_{+}^2\right] -g_a^* g_b h -g_a g_b^* h^* -  {|h|}^2 \left(\omega-\Omega+ i\frac{1}{\tau_q}\right)},\label{E:th}\\
r &=\frac{ -i \frac{V_a^* V_b}{v_g}\left[g_a g_b^*+ h \left(\omega-\Omega+ i\frac{1}{\tau_q}\right)\right]}{ \left(\omega -{\omega_c}+ i \frac{1}{\tau_c}+i \Gamma\right)  \left[\left(\omega-\Omega+ i\frac{1}{\tau_q}\right) \left(\omega -{\omega_c}+ i \frac{1}{\tau_c}+i \Gamma\right)-G_{+}^2\right] -g_a^* g_b h -g_a g_b^* h^* -  {|h|}^2 \left(\omega-\Omega+ i\frac{1}{\tau_q}\right)},\label{E:rh}\\
e_a &=\frac{V_a^*\left[\left(\omega-\Omega+i\frac{1}{\tau_q}\right)\left(\omega-\omega_c+i\frac{1}{\tau_c}+i \Gamma\right)-{|g_b|}^2\right]}{ \left(\omega -{\omega_c}+ i \frac{1}{\tau_c}+i \Gamma\right)  \left[\left(\omega-\Omega+ i\frac{1}{\tau_q}\right) \left(\omega -{\omega_c}+ i \frac{1}{\tau_c}+i \Gamma\right)-G_{+}^2\right] -g_a^* g_b h -g_a g_b^* h^* -  {|h|}^2 \left(\omega-\Omega+ i\frac{1}{\tau_q}\right)}\label{E:eah},\\
e_b &=\frac{V_a^*\left[g_a g_b^*+h \left(\omega-\Omega+i\frac{1}{\tau_q}\right)\right]}{ \left(\omega -{\omega_c}+ i \frac{1}{\tau_c}+i \Gamma\right)  \left[\left(\omega-\Omega+ i\frac{1}{\tau_q}\right) \left(\omega -{\omega_c}+ i \frac{1}{\tau_c}+i \Gamma\right)-G_{+}^2\right] -g_a^* g_b h -g_a g_b^* h^* -  {|h|}^2 \left(\omega-\Omega+ i\frac{1}{\tau_q}\right)}\label{E:ebh},\\
e_q &=\frac{V_a^*\left[g_a\left(\omega-\omega_c + i\frac{1}{\tau_c}+i \Gamma\right)+h g_b\right]}{ \left(\omega -{\omega_c}+ i \frac{1}{\tau_c}+i \Gamma\right)  \left[\left(\omega-\Omega+ i\frac{1}{\tau_q}\right) \left(\omega -{\omega_c}+ i \frac{1}{\tau_c}+i \Gamma\right)-G_{+}^2\right] -g_a^* g_b h -g_a g_b^* h^* -  {|h|}^2 \left(\omega-\Omega+ i\frac{1}{\tau_q}\right)}\label{E:eqh},
\end{align}
\end{subequations}
}where we have used $|V_a|=|V_b| \equiv V$, $\Gamma \equiv \frac{V^2}{2 v_g}$, $G_{+}^2\equiv {|g_a|}^2+{|g_b|}^2$, and $G_{-}^2\equiv {|g_b|}^2-{|g_a|}^2$, all are real numbers. Notice that $\Gamma$ is the external linewidth of the WGM's due to waveguide-cavity coupling. These analytic expressions of the amplitudes, Eq.~(\ref{E:th})-(\ref{E:eqh}), provide a complete description on the single-photon transport properties. These equations are applicable to arbitrary two-level systems having any orientation of the electric dipole moment, specified through $g_a$ and $g_b$.

\section{General Properties of The Transmission Spectrum}\label{Sec:GeneralProperties}

The expressions of the amplitudes (Eqs.~(\ref{E:Amplitudes})) allow us to make exact statements on the general properties of the spectrum of the full coupled system, such as the spectral symmetry properties and the generalized critical coupling conditions. These statements are valid even in the presence of resonator and atom dissipations. We will exemplify these general properties with concrete examples in the following sections.

\subsection{Spectral Symmetry Properties of The Transmission Spectrum}

It is straightforward to show that the transmission amplitude of Eq.~\eqref{E:th} satisfies the following spectral symmetry condition, \emph{regardless of the values of the cavity and atom dissipations}:
\begin{equation}\label{E:General}
t^*\left(-\delta\omega,  -\Delta, -\delta\theta\right) = t(\delta\omega,  \Delta, \delta\theta),
\end{equation}where $\delta\omega\equiv \omega-\omega_c$,  $\Delta\equiv\Omega-\omega_c$ is the frequency detuning between the resonator and the atom, and $\delta\theta\equiv \theta_h+\theta_b-\theta_a - \frac{\pi}{2}$ is the phase mismatch. This condition states that the transmission spectrum $T(\omega)\equiv {|t(\omega)|}^2$ of one system with parameters $(-\Delta, -\delta\theta)$ is mirror-imaged with respect to $\omega_c$ to that of another system with parameters $(\Delta, \delta\theta)$.

In particular, for a system with $\Delta=0$ and $\delta\theta\equiv 0\,\, (\text{mod}\,\, 2\pi)$, the transmission spectrum is symmetric with respect to $\omega=\omega_c$: $T(\omega=\omega_c+\delta\omega)=T(\omega=\omega_c-\delta\omega)$. The former condition $\Delta=0$ says the resonator and the atom are in-tuned ($\Omega=\omega_c$); while the latter condition $\delta\theta\equiv 0\,\, (\text{mod}\,\, 2\pi)$ requires $g_a^* g_b h$ to be purely imaginary so that $g_a^* g_b h +g_a g_b^* h^*=0$. The same spectral symmetry property holds for other amplitudes of Eq.~(\ref{E:rh})-(\ref{E:eqh}). 

%When these two conditions are satisfied, the spectrum is always symmetric with respect to $\omega=\omega_c$, regardless to the values of all the parameters. Note the phase dependence of the amplitudes upon $g$'s and $h$ is through the phase of $g_a^* g_b h$, \emph{i.e.}, $\theta_h+\theta_b-\theta_a$. Thus, only the relative phase between these coupling constants are important.In the following discussions, without losing generality, we assume $g$ is real, while $h$ is complex.

\subsection{Generalized Critical Coupling Condition}

Critical coupling is defined for a system of coupled waveguides and resonators when transmission of the input signal goes to zero at the output port at resonance. The existence of critical coupling is of direct interest to quantum optics experiments where one often desires to eliminate single-photon transmission so that the signatures of two-photon transmission are distinct from single-photon effects. Here we present the condition under which critical coupling can be achieved for single-photon transport in the full coupled waveguide-ring resonator-atom system.

By requiring the transmission amplitude of Eq.~\eqref{E:th} to be zero, one can show that the critical coupling is reached at $\omega=\omega_c$ if the following criterions are satisfied:
\begin{enumerate}
\item The resonator and the atom are in-tuned, \emph{i.e.,} $\Omega-\omega_c \equiv \Delta =0$; and
\item 
\begin{subequations}
\begin{align}
\Gamma^2 + \tau_q G_{-}^2\Gamma &= \left(\frac{1}{\tau_c}\right)^2 + \frac{\tau_q}{\tau_c} G_{+}^2 + |h|^2,\label{E:Generalize}\\
g^*_a g_b h + g_a g^*_b h^* &=0.\label{E:GeneralizePhase}
\end{align}
\end{subequations}
Eq.~\eqref{E:Generalize} is a magnitude-matching condition. Eq.~\eqref{E:GeneralizePhase} is a phase-matching condition, requiring $g^*_a g_b h$ to be purely imaginary. Since $g^*_a g_b h = i |g_a| |g_b| |h| e^{i \delta\theta}$, with $\delta\theta \equiv\theta_h + \theta_b -\theta_a -\pi/2$, Eq.~\eqref{E:GeneralizePhase} is equivalent to the phase-matching condition $\delta\theta = 0 \,(\text{mod}\,\, 2\pi)$.
\end{enumerate}
%The generalized critical coupling conditions therefore consist of a pair of conditions: the magnitude matching condition of Eq.~\eqref{E:Generalize}, and the phase matching condition of Eq.~\eqref{E:GeneralizePhase}.

%\begin{equation}
%\Gamma^2 + \tau_q G_{-}^2\Gamma = \left(\frac{1}{\tau_c}\right)^2 + \frac{\tau_q}{\tau_c} G_{+}^2 + |h|^2.
%\end{equation}This is the generalized critical coupling condition. 

%We will illustrate these general properties with concrete examples in later sections.

\section{Numerical Validation}\label{Sec:Numerical}

Before we proceed to understand the detailed predictions of Eqs.~\eqref{E:Amplitudes},  here we first compare Eq.~(\ref{E:th})-(\ref{E:eqh}) to experimental data of waveguide-resonator systems~\cite{Aoki:2006, Kippenberg:2004}, and to numerical results of Srinivasan and Painter's~\cite{Srinivasan:2007} on the waveguide-microdisk-quantum dot system. In all cases the agreements are excellent.

\subsection{Fitting to Experimental Data of Waveguide-Resonator Systems}\label{Sec:NumericalWR}

\begin{figure}[htb]
%\scalebox{1}{\includegraphics{WRNosc.eps}}
\scalebox{1}{\includegraphics[width=\columnwidth]{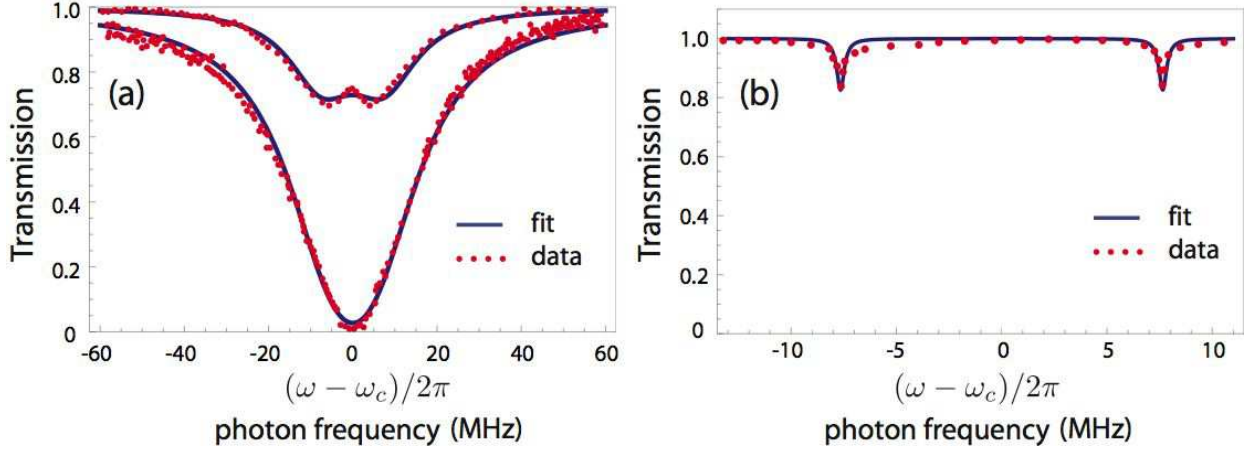}}
\caption{(Color online) Fitting to experimental data of system of waveguide-microtoroidal resonator. $\Gamma\equiv {|V|}^2/2v_g$. (a) Lower curve: $|h|/2\pi=5.50092$ MHz, $1/\tau_c/2\pi=7.49983$ MHz. $\Gamma/2\pi= 6.89887$ MHz. $\Gamma^2=1.34818^2 \sqrt{{|h|}^2+(1/\tau_c)^2}$, which is somewhat close to the critical coupling condition.  Upper curve: under-coupling. Blue curve: $|h|/2\pi=7.57069$ MHz, $1/\tau_c/2\pi=8.4642$ MHz. $\Gamma/2\pi=1.13735$ MHz.  $\Gamma^2 \ll {|h|}^2+(1/\tau_c)^2$. The overall decay rate $\kappa$ in Ref.~[\onlinecite{Aoki:2006}] is equal to $\Gamma+1/\tau_c$ in this article. (b) $|h|/2\pi=7.64947$ MHz, $1/\tau_c/2\pi=0.250879$ MHz, $\Gamma/2\pi = 0.0194301$ MHz. $\Gamma^2 \ll {|h|}^2+(1/\tau_c)^2$ and $|h| \gg 1/\tau_c \gg \Gamma$. The numerical values are obtained by least squares method using the amplitude of Eq.~\eqref{E:WRt2}.}\label{Fi:Fitting}
\end{figure}

To demonstrate the validity of the exact solutions, we apply these expressions to fit two of the recent experimental data of waveguide-microtoroidal resonator systems~\cite{Aoki:2006, Kippenberg:2004}. The results are plotted in Fig.~(\ref{Fi:Fitting}). The amplitudes for the waveguide-resonator system can be deduced from those of the full coupled waveguide case by decoupling the atom. In particular, the transmission amplitude is given by
\begin{equation}\label{E:WRt2}
t(\omega)=\frac{\left(\omega-\omega_c+i\frac{1}{\tau_c}\right)^2-{|h|}^2+\Gamma^2}{\left(\omega-\omega_c+i\frac{1}{\tau_c}+i\Gamma\right)^2-{|h|}^2}.
\end{equation}The details are given in Appendix~\ref{A:List}. Note only the magnitude of $h$ appears in the amplitude, its phase does not.

Fig.~(\ref{Fi:Fitting})(a) shows transmission spectra of a waveguide-microtoroidal resonator system~\cite{Aoki:2006}. In the experiments, the waveguide-resonator coupling strength $|V|$ is varied by adjusting the distance between the waveguide and the microtoroid. The lower trace is taken such that the critical coupling condition is approximately satisfied: $\Gamma^2 \simeq {|h|}^2+(1/\tau_c)^2$; while the upper trace is for conditions of under-coupling: $\Gamma^2 \ll {|h|}^2+(1/\tau_c)^2$.   By fitting to the spectra using Eq.~\eqref{E:WRt2}, one can extract the numerical values of $|h|$, $1/\tau_c$, and $\Gamma$. Our numerical results show that $|h| \simeq 1/\tau_c$.

When the same values of $|h|$ and $1/\tau_c$ determined from fitting to the lower trace are used to fit the upper curve, the width of the spectrum is narrower than that of the experimental data, and the minimum region has significant deviation. If, however, somewhat larger values of $|h|$ and $1/\tau_c$ are used instead to fit the data, the spectrum can be fitted reasonably well, as shown by the upper blue curve in Fig.~(\ref{Fi:Fitting})(a). This indicates that both the inter-mode backscattering and the intrinsic loss of the resonator are slightly suppressed when the coupling between the waveguide and the resonator is strong.

As another validation, in Fig.~(\ref{Fi:Fitting})(b), we use Eq.~\eqref{E:WRt2} to fit the transmission spectrum of another  waveguide-microtoroidal resonator system~\cite{Kippenberg:2004}, which is in the deep under-coupling regime such that $\Gamma^2 \ll {|h|}^2 + \left(1/\tau_c\right)^2$. The numerical fitting indicates that $|h| \gg 1/\tau_c \gg \Gamma$ so the inter-mode backscattering dominates. In this regime, the transmission spectrum shows a doublet structure. The doublet structure in the transmission spectrum induced by the inter-mode backscattering for ultra-high-Q resonator in the $|h| > \Gamma$ regime is well-known~\cite{Weiss:1995, Kippenberg:2002,Kippenberg:2004}.

In both cases above, the extracted numerical values agree reasonably well with those reported in the experiments.

\subsection{Comparing to Numerical Results of Waveguide-Resonator-Atom Systems}\label{Sec:NumericalWRA}

We next compare our analytic results of the amplitudes of Eqs.~\eqref{E:th}-\eqref{E:eqh} to the numerical results on coupled waveguide-microdisk-quantum dot system in Srinivasan and Painter~\cite{Srinivasan:2007}, where $g_b = g_a$ by properly choosing the orientation of the quantum dot dipole polarization, and the azimuthal origin. From Eqs.~\eqref{E:th}-\eqref{E:eqh}, one immediately sees that all amplitudes, except $e_q$, depend on the magnitude of $g_a$ only but not its phase. The atom excitation amplitude $e_q$ is  proportional to $g_a$, the spectrum ${|e_q|}^2$ however also depends on $|g_a|$ only.

\begin{figure}[htb]
%\scalebox{0.6}{\includegraphics{FittingQME.eps}}
\scalebox{1}{\includegraphics[width=\columnwidth]{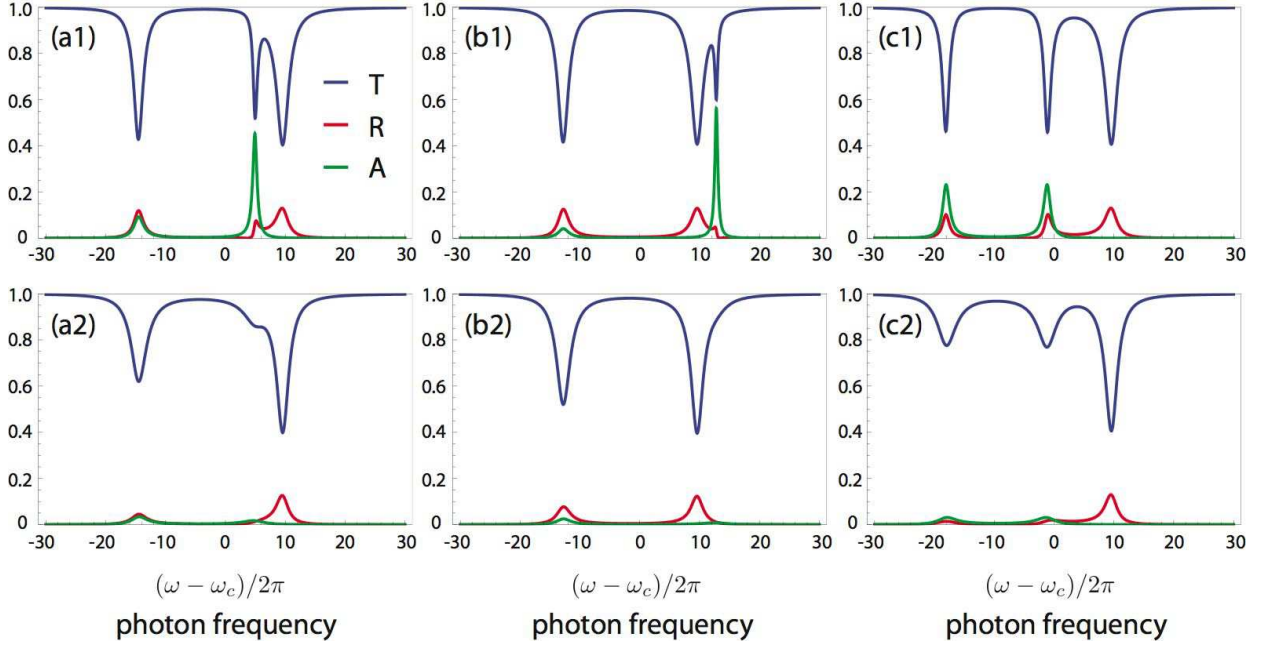}}
\caption{(Color online) The transmission spectrum $T$ (blue curve), reflection spectrum $R$ (red curve), and the atom excitation spectrum $A\equiv {|e_q|}^2$ (green curve) of the coupled waveguide-microdisk-quantum dot system. The normalized spectrum $\frac{\Gamma}{2 v_g} {|e_q|}^2$ is plotted. Upper panel ((a1)-(c1)): dephasing $\gamma_p=0$. Lower panel ((a2)-(c2)): $\gamma_p\neq0$. (a1), (a2): $\Omega=\omega_c$. (b1), (b2):  $\Omega-\omega_c = |h|$. (c1), (c2): $\Omega-\omega_c = -|h|$. Parameters used: $g/2\pi = 6$, $h/2\pi=-9.6$,  $1/\tau_q/2\pi=0.16$, $1/\tau_c/2\pi= 0.76$, $\Gamma/2\pi=0.44$, and $\gamma_p/2\pi=2.4$.}\label{Fi:FittingQME}
\end{figure}

Fig.~\ref{Fi:FittingQME} plots the transmission and reflection spectra
%for the case $h$ is real, 
using the parameters corresponding to those in Fig.~5 of Ref.~[\onlinecite{Srinivasan:2007}]. (The relation between the parameters used in this article and Ref.~[\onlinecite{Srinivasan:2007}] is: $g=|g_a|=|-i g_0|$, $\Gamma=\kappa_e$, $1/\tau_c = \kappa_i=\kappa_T - \kappa_e$, $1/\tau_q=\gamma_{\|} + \gamma_p$, and $h=-\beta$). 
In Fig.~\ref{Fi:FittingQME}, the upper panel plots the spectra when the quantum-dot dephasing $\gamma_p$ is zero; while the lower panel plots the spectra when $\gamma_p \neq 0$. Each panel plots the in-tuned ($\Omega=\omega_c$) and de-tuned ($\Omega \neq \omega_c$) cases. The results of both approaches are in excellent agreement. We were able to obtain excellent agreements with all results in Ref.~[\onlinecite{Srinivasan:2007}] (not shown here).

In Fig.~\ref{Fi:FittingQME} we also plot the atom excitation. 
%The atom excitation is strongly suppressed by the dephasing, as can be seen when comparing the lower and upper panels of Fig.~\ref{Fi:FittingQME}.  
In general, there are only two resonances in the atom excitation, in contrast to the three resonances in the transmission spectrum. Since the atom is an intermediary for correlating photons, one expects the photon-photon correlations are qualitatively different at the three resonances of the transmission spectrum. This is supported by comparing the atomic excitation spectra with the photon correlation function $g^{(2)}(\tau)$ in Fig.~12 of Ref.~[\onlinecite{Srinivasan:2007}]. 

% The degree of the atomic excitation gives strong indication to the photon-photon correlation function, \emph{i.e.,} the $g^{(2)}$ function, since the atom is an intermediary for correlating photons. This can be clearly seen by comparing the atomic excitation spectrum in Fig.~\ref{Fi:FittingQME}(c2) with the $g^{(2)}(\tau)$  in Fig.~12 of Srinivasan and Painter. The atom is slightly excited at the two lower resonance frequencies of the reflection spectrum in Fig.~\ref{Fi:FittingQME}(c2), but is essentially unexcited at the highest resonance frequency, resulting an uncorrelated $g^{(2)}(\tau)$. In Fig.~12(c) of Ref.~\cite{SrinivasanParameter}, one can see that the $g^{(2)}(\tau)$ is essentially unity for all times. Note that reducing the atom dissipation or dephasing does not change the degree of photon-photon correlation at this frequency. Fig.~\ref{Fi:FittingQME}(c1) plots these amplitudes with the atom dephasing being zero, where one can see that the atom is still unexcited. This is due to the cavity nature at this resonance frequency. If the resonance peak is of an atomic nature, atomic dissipation and dephasing will have crucial effects on the atom excitation, and in turn, on the photon correlations, as seen in Fig.~\ref{Fi:FittingQME}(a1)(a2), and (b1)(b2). Below we will also show that the atom excitation can only have at most two resonance peaks. Other figures in Ref.~\cite{SrinivasanParameter} could be plotted using Eqs.~\eqref{E:th}-\eqref{E:eqh} straightforwardly and again show excellent agreement between these two approaches (not shown here).

\section{Single-Photon Transport of The Coupled Waveguide-Resonator-Atom Systems}\label{Sec:WRA}

We now turn the attentions to the single-photon transport of the full coupled waveguide-resonator-atom system and the system responses. The analytic results enable us to present a systematic parametric study of the transport properties by continuously varying the parameters. Such a systematic study facilitates the understanding of the underlying physics in this complicated coupled systems. As a concrete example, we will take $g_a = g_b$, and let $g\equiv |g_a|$ throughout the discussions hereafter. This choice of coupling constants correspond to that considered in Ref.~[\onlinecite{Srinivasan:2007}]. Other choices of the coupling constants could be investigated in the same manner.

In this section the inter-mode backscattering strength $h$ is taken as real and the system is assumed lossless ($1/\tau_c=1/\tau_q=0$).  The atom and the resonator are also assumed to be in-tune ($\Omega=\omega_c$). The effect of dissipation, complex $h$, and atom-cavity detuning will be considered in later sections.

We consider the spectra of transmission $T\equiv {|t|}^2$ (Fig.~\ref{Fi:TArray}), the group delay $d\phi/d\omega$ (Fig.~\ref{Fi:DelayArray}), the atom excitation ${|e_q|}^2$ (Fig.~\ref{Fi:EqArray}), and the phase matched WGM excitation  ${|e_a|}^2$ (Fig.~\ref{Fi:EaArray}) for different value of $g$ and $h$. (For completeness, we also plot the spectra of reflection $R\equiv |r(\omega)|^2$ and the counter-propagating WGM excitation $|e_b|^2$ in Appendix~\ref{A:ReflectionAndWGM}). To facilitate visualization, these spectra are presented in a matrix form with different rows or columns corresponding to different $g$ and $h$, respectively. Comparing this set of spectra provides very useful information to understand the full coupled system, since the atom and the WGM excitation determine the nature of the transmission resonances. Below, we refer to a resonance with large (small) atom excitation as a resonance of atom (cavity) nature.  
%The characterization of the nature of a resonance is crucially important for understanding the effects of the dissipation: for a resonance of atom nature, for example, the resonance is strongly affected by the atom dissipation, but is only weakly affected by the cavity dissipation. For a resonance of pure atom nature, the resonance could even be insensitive to the cavity dissipation. Similar scenario occurs for a resonance of cavity nature.

%In this section $h$ is taken as real and the system is assumed lossless ($1/\tau_c=1/\tau_q=0$).  The dissipative and the complex $h$ cases will be discussed in the following sections. The atom and the resonator are also assumed to be in-tune ($\Omega=\omega_c$) for now.

%We now discuss these visualization maps, and describe the qualitative features of these maps. The quantitative features which substantiate the qualitative descriptions will be given in the later sections.

\subsection{First column: the atom decoupled}

The leftmost column describes the waveguide-resonator case with the atom decoupled:
%The transmission amplitude of the coupled waveguide-resonator is given by
%\begin{equation}\label{E:WRt}
%t(\omega)=\frac{\left(\omega-\omega_c+i\frac{1}{\tau_c}\right)^2-{|h|}^2+\Gamma^2}{\left(\omega-\omega_c+i\frac{1}{\tau_c}+i\Gamma\right)^2-{|h|}^2},
%\end{equation}with $\Gamma\equiv \frac{{|V|}^2}{2 v_g}$. This expression can be easily obtained by either starting with the Hamiltonian $H$ in Eq.~(\ref{E:Hamiltonian}) and the state $|\epsilon^{+}\rangle$ in Eq.~(\ref{E:InteractingEigenstate}) by omitting the atomic parts; or from the expression of $t$ in Eq.~(\ref{E:th}) by letting the atomic transition frequency $\Omega\rightarrow\infty$ and the coupling $g\rightarrow 0$, which effectively decouples the atom from the waveguide-resonator part.  Note that when the atom is decoupled, the transmission amplitude $t$ is a function of the magnitude of the mode cross-talking strength, $|h|$, only, but not its phase.
\begin{enumerate}
\item When $h=0$, the resonator acts as an all-pass filter with transmission $T= 1$ for all frequency. The group delay shows a Lorentzian peak.
\item When $h$ is slightly increased but small than $\Gamma$, a transmission dip develops at the resonance frequency $\omega=\omega_c$, with
\begin{equation}
0 < T(\omega_c)=\left(\frac{{|h|}^2-\Gamma^2}{{|h|}^2+\Gamma^2}\right)^2<1.
\end{equation}Also, the transmission has a non-Lorentzian lineshape.
\item When $h=\Gamma$, the transmission spectrum has a flat bottom centered at $\omega=\omega_c$ where the transmission becomes zero. The transmission has a maximally-flat 2nd-order Butterworth filter lineshape given by
\begin{equation}
T(\omega)=\frac{\left(\omega-\omega_c\right)^4}{\left(\omega-\omega_c\right)^4 + 4{|h|}^4},
\end{equation}with a full width at half maximum (FWHM) equal to $2\sqrt{2}|h|$. The group delay however shows two splitted peaks.
\item When $h>\Gamma$, the transmission shows two resonance dips at which the transmission is zero. The spectral separation between the two dips is $2\sqrt{|h|^2-\Gamma^2}$, which approaches $2|h|$ when $|h| \gg \Gamma$. 
\end{enumerate}

Thus the qualitative behavior of the transmission spectrum is determined by the ratio of ${|h|}/\Gamma$. For an ultra-high $Q$ resonator with very weak external coupling to the waveguide, \emph{i.e.}, very small value of $\Gamma$, even a weak mode cross-talking would qualitatively change the transmission spectrum. 

Since the atom is decoupled, all the transmission dips(resonances) have a cavity nature, \emph{i.e.}, the atom excitation is zero.

\subsection{First row: no inter-mode backscattering in the resonator}

The uppermost row describes an ideal resonator without mode cross-talking, \emph{i.e.}, $h=0$:
\begin{enumerate}
\item When $g=0$, the atom is decoupled, and the resonator is an all-pass filter as discussed above.
\item When $g$ is slightly increased from zero such that $2 g^2 \ll \Gamma^2$ (shown in figure: $g=0.3\Gamma$), a transmission dip down to zero occurs at frequency $\omega=\omega_c$. This transmission dip is induced by the atomic resonance and is in contrast to that induced by small $h$ above, which does not go to zero. In this regime, the poles of the transmission amplitude are
$\omega-\omega_c = -i\frac{2g^2}{\Gamma}$, $-i (\Gamma -\frac{2g^2}{\Gamma})$, and $-i\Gamma$. The transmission spectrum can be well approximated by a Lorentzian with FWHM equal to $4 g^2/\Gamma$. Small deviation of the spectrum from a Lorentzian occurs only when $|\omega-\omega_c|\simeq \Gamma$, where the poles with larger imaginary parts become important in the spectral response. Moreover, the atom excitation has a strong peak at $\omega=\omega_c$, thus the transmission dip has an atomic nature. The group delay shows a single peak.  Notice that the group delay in this case is much larger than case B1 above.
\item When $g$ is further increased such that $2 g^2 = \Gamma^2$, the transmission spectrum has a flat bottom centered at $\omega=\omega_c$ where the transmission is zero. The transmission has a maximally-flat 3rd-order Butterworth filter lineshape given by
\begin{equation}
T(\omega) = \frac{(\omega-\omega_c)^6}{(\omega-\omega_c)^6+8|g|^6},
\end{equation}
with a full width at half maximum (FWHM) equal to $2\sqrt{2}g$. Both the group delay and the atom excitation however shows two splitted peaks. 
\item When $g$ is further increased such that $2 g^2 > \Gamma^2$ (shown in figure: $g=2.5\Gamma$), three transmission dips develop. This is because the two counter-propagating WGMs of the resonator, through linear superposition, can alternatively be described as  two standing-wave modes. One of the standing-wave modes has zero amplitude at the atom location, and corresponds to the middle transmission dip at $\omega=\omega_c$. This resonance is of cavity nature. The other mode has non-zero amplitude at the atom location, and experiences Rabi-splitting through interacting with the atom. The transmission spectra exhibits three dips when such Rabi splitting is large enough compared with $\Gamma$. The two side dips, which result from the Rabi splitting, have a mixture of cavity and atom nature, as confirmed by examining the atom excitation plot. The spectral separation between the two side dips is $2\sqrt{2 {|g|}^2 -\Gamma^2}$. In addition, the two side-dips have a larger group delay than that of the middle peak, since the photon experiences delay from both the cavity and the atom.
\item When  $2 g^2 \gg \Gamma^2$ (shown in figure: $g=8\Gamma$), the spectral separation between the two dips approaches $2\sqrt{2} |g|$. The maximum transmission  between each side-dip and the middle dip approaches 1 as $1-\frac{27}{4}\frac{\Gamma^2}{2{|g|}^2}$. Moreover, the width of the middle dip is the sum of the widths of the two side-dips.
\end{enumerate}

The case of $h=0$ represents the ideal case for WGM resonators and therefore is of fundamental importance. In Appendix~\ref{A:hZero}, we summarize all analytic results related to this important case.

%To gain some insights into these complicated analytic expressions, we will start by investigating some simpler cases and then gradually build up the complexities. In the following, we thus will first discuss the waveguide-ring resonator system, followed by discussing the full waveguide-resonator-atom system. For each case, we will discuss the effects of the cavity and atom dissipations.

\subsection{General case: non-zero $h$ and $g$}

\begin{figure}[htb]
%\scalebox{1.4}{\includegraphics{T_Array.eps}}
\scalebox{1}{\includegraphics[width=\columnwidth]{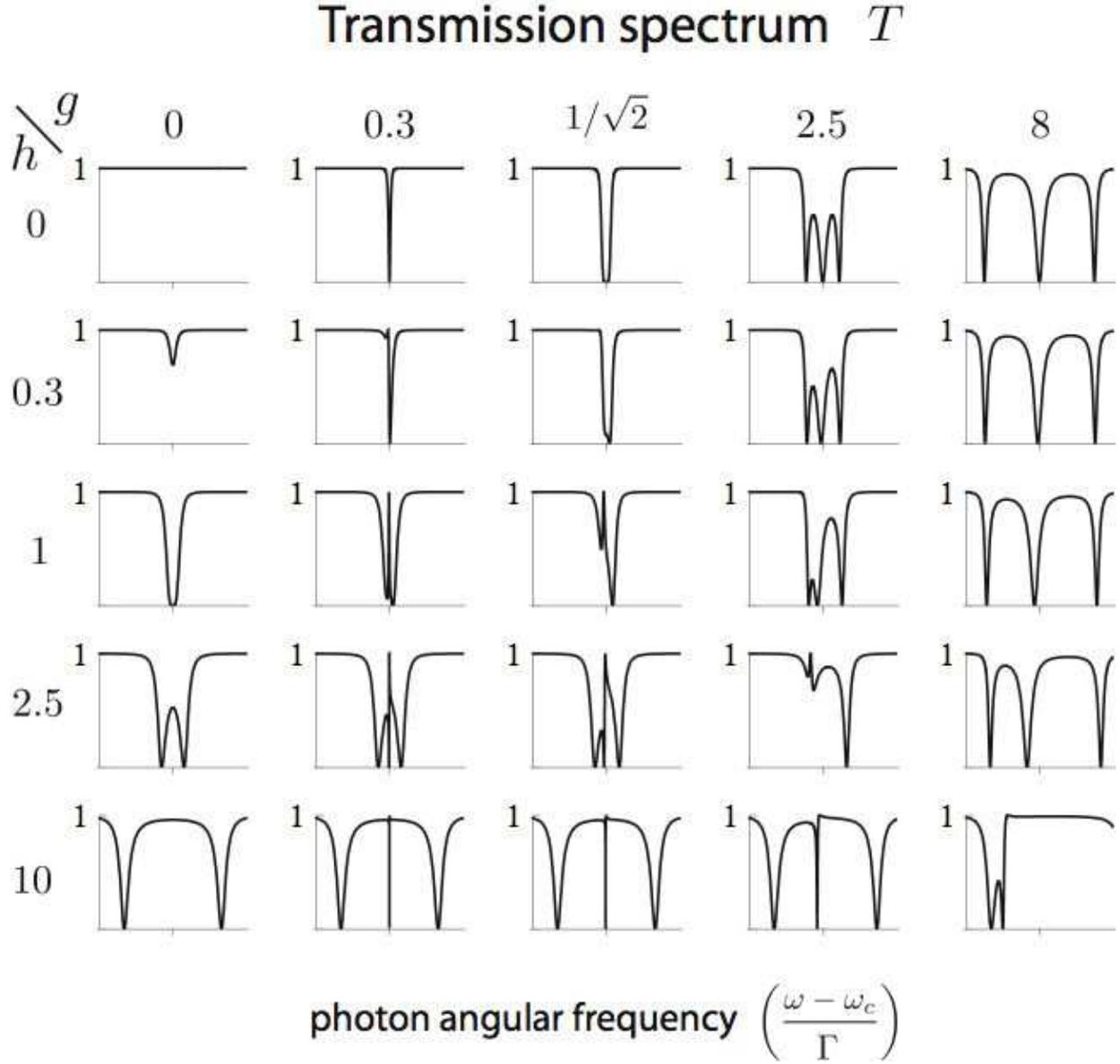}}
\caption{Map of the transmission spectrum $T$ as $g$ and $h$ are varied. $h$ is real for this map. Also, $g_a = g_b$ and $g\equiv |g_a|$. The location of $\omega=\omega_c$ is indicated by the gray tick in each figure.}\label{Fi:TArray}
\end{figure}

\begin{figure}[htb]
%\scalebox{1.4}{\includegraphics{Delay_Array.eps}}
\scalebox{1}{\includegraphics[width=\columnwidth]{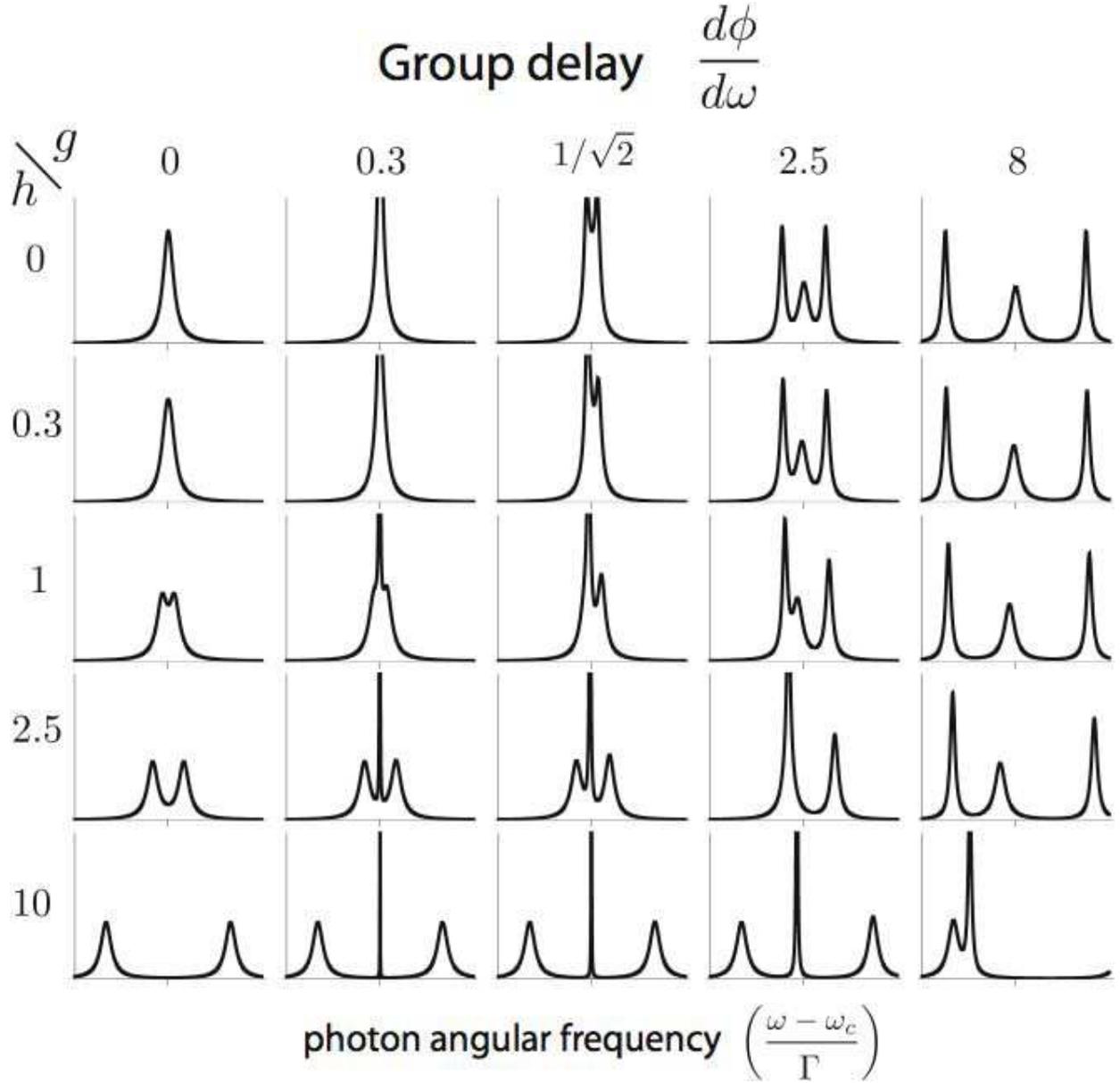}}
\caption{Map of the group delay $d\phi/d\omega$ as $g$ and $h$ are varied. $h$ is real for this map. Also, $g_a = g_b$ and $g\equiv |g_a|$. The location of $\omega=\omega_c$ is indicated by the gray tick in each figure.}\label{Fi:DelayArray}
\end{figure}

\begin{figure}[htb]
%\scalebox{1.4}{\includegraphics{Eq_Array.eps}}
\scalebox{1}{\includegraphics[width=\columnwidth]{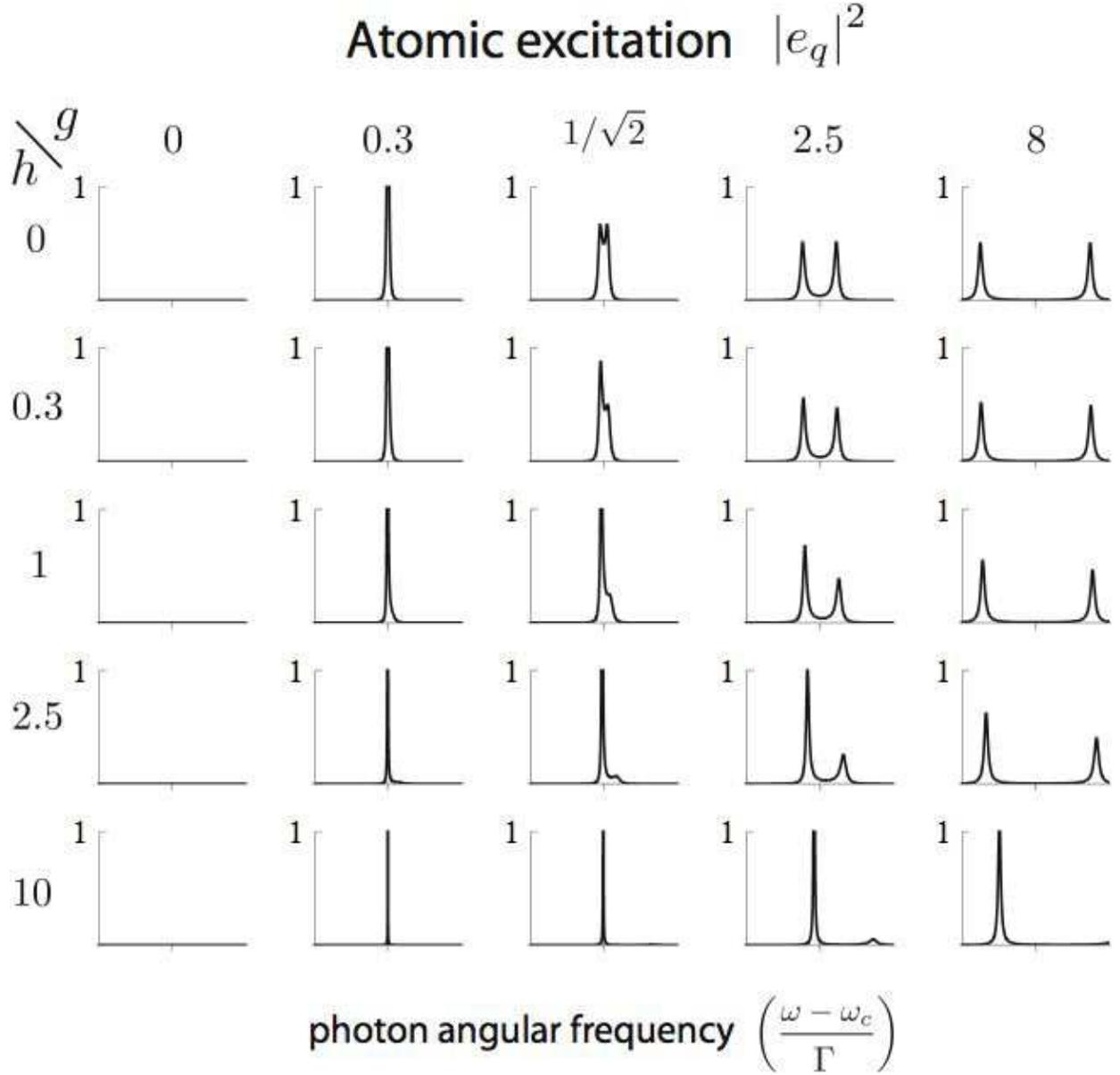}}
\caption{Map of the atomic excitation spectrum ${|e_q|}^2$ as $g$ and $h$ are varied. $h$ is real for this map. Also, $g_a = g_b$ and $g\equiv |g_a|$. The normalized spectrum $\frac{\Gamma}{2 v_g} {|e_q|}^2$ is plotted. The location of $\omega=\omega_c$ is indicated by the gray tick in each figure.}\label{Fi:EqArray}
\end{figure}

\begin{figure}[htb]
%\scalebox{1.4}{\includegraphics{Ea_Array.eps}}
\scalebox{1}{\includegraphics[width=\columnwidth]{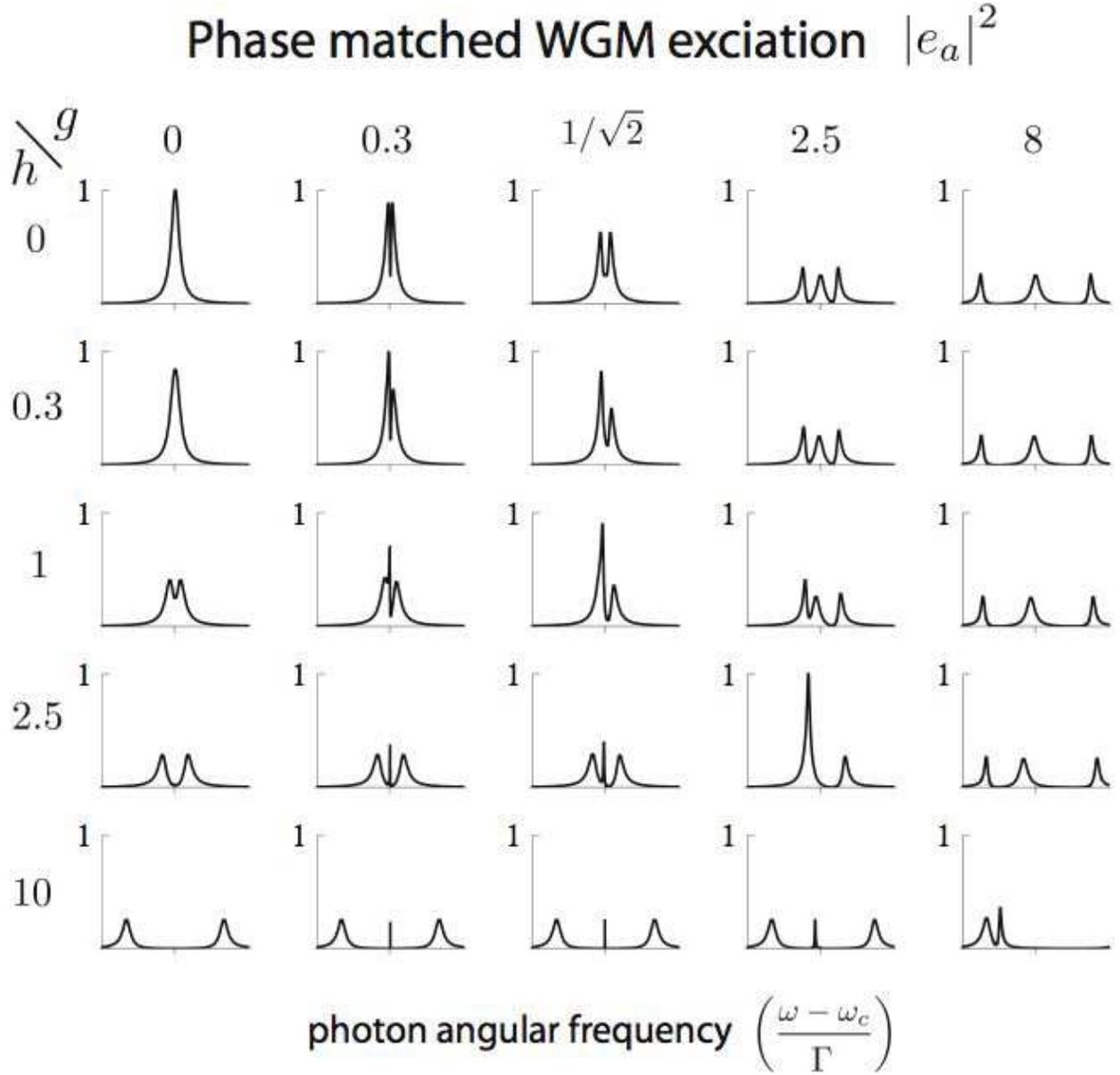}}
\caption{Map of the phase-matched WGM excitation ${|e_a|}^2$ as $g$ and $h$ are varied. $h$ is real for this map. Also, $g_a = g_b$ and $g\equiv |g_a|$. The normalized spectrum $\frac{\Gamma}{2 v_g} {|e_a|}^2$ is plotted. The location of $\omega=\omega_c$ is indicated by the gray tick in each figure.}\label{Fi:EaArray}
\end{figure}

\begin{figure}[htb]
\scalebox{1}{\includegraphics{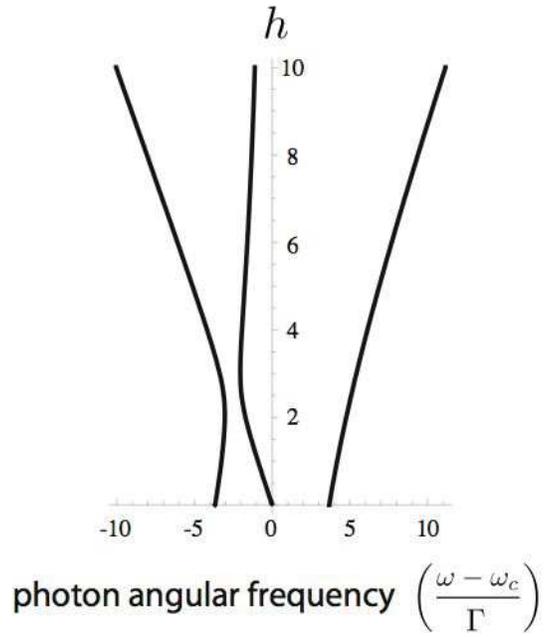}}
%\scalebox{1}{\includegraphics[width=\columnwidth]{AntiCrossing.eps}}
\caption{Anti-crossing between the atomic and cavity resonances as $h$ is continuously varied. $g=2.5 \Gamma$. This depicts the evolution of the resonances in the column of $g=2.5 \Gamma$ in Fig.~\ref{Fi:TArray}. %Right: $g$ is varied. $h=2.5 \Gamma$. This depicts the evolution of the resonances in the row of $h=2.5 \Gamma$ in Fig.~\ref{Fi:TArray}
}\label{Fi:AntiCrossing}
\end{figure}

We now discuss the general features of the cases when both $h$ and $g$ are non-zero. 
\begin{enumerate}
\item Comparing the leftmost two columns in Fig.~\ref{Fi:TArray}, where we increase $g$ slightly from zero to a small non-zero value ($0.3\Gamma$ in the figures). For all values of $h$, the effect of introducing an atomic resonance by a small $g$ is to create a resonance of atomic nature at $\omega=\omega_c$, resulting in an asymmetrical Fano-type lineshape in transmission~\cite{Fan:2003}, and a large group delay at resonance. The rest of the spectrum is not significantly perturbed. 
\item Comparing the uppermost two rows in Fig.~\ref{Fi:TArray}, where we increase $h$ slightly from zero to a small non-zero value ($0.3\Gamma$ in the figures). For all values of $g\neq 0$, the effect of introducing an inter-mode backscattering by $h$ is to slightly distort the transmission spectrum to be asymmetrical.
\item  The cases with intermediate values of $g$ and $h$ can be obtained from perturbing from its counterpart in each direction. The exact lineshapes have to be computed using the analytic expressions.
\item One important feature is the anti-crossing between the atomic and cavity resonances as one varies the values of $h$ or $g$. We will demonstrate this point using the column of $g=2.5\Gamma$. Fig.~\ref{Fi:AntiCrossing} plots the real part of the poles of the transmission amplitude $t(\omega)$, which indicate the spectral locations of the resonances. 
%When $h \simeq 2$ to $4\,\Gamma$, the left two resonances anti-cross, and exchange the cavity or atom nature of the resonances. Thus, for the column of $g=2.5\Gamma$, a
At $h=0$, the transmission spectrum starts with three resonances with the middle dip of  cavity nature, and the two side-dips mixture of both cavity and atom nature, as previously explained. When $h$ is increased (from $h=0.3\Gamma$ to $10\,\Gamma$), the left two dips anticross. During the anti-crossing process, the nature of the resonance of the two dips is exchanged, as can be seen by examining the weight of the atomic excitation along the evolution process. In the end, when $h$ becomes $10\Gamma$, the left resonance is of purely cavity nature, while the middle resonance is of purely atomic nature, and the right resonance is of largely cavity nature with a little atomic mixture. Similar anti-crossing behavior can be observed when $g$ is continuously varied instead. Thus, it is difficult to ascertain the nature of a resonance judging from the position of the resonance alone, in the case where significant back scattering occurs.

%Similarly, for the row of $h=2.5\Gamma$, when $g$ is changed from $0$ to $8\Gamma$, the transmission spectrum starts with two resonance of purely cavity nature ($g=0$), to three resonances with the middle of purely atomic nature while the two side-dips of pure cavity nature. As $g$ increases from $0.3\Gamma$ to $2.5\Gamma$, a second atomic resonance develops on the right. When $g=8 \Gamma$, the transmission spectrum has three resonances, with the middle resonance of purely cavity nature, while the two side-resonances of mixture of cavity and atomic nature. 
\end{enumerate}

\section{Effects of Dissipations}\label{Sec:Dissipations}

The effects of intrinsic dissipations on the transmission resonances strongly depend upon the nature of the resonances. When a resonance is of purely cavity nature, the transmission at the resonance frequency is insensitive to the atom dissipation, but only to the cavity dissipation. Similarly, when a resonance is of purely atomic nature, the transmission at the resonance frequency is insensitive to the cavity dissipation, but only to the atom dissipation. For a resonance of mixed nature, the transmission is affected by either type of dissipation. 

\begin{figure}[htb]
%\scalebox{1.4}{\includegraphics{T_Array_Diss.eps}}
\scalebox{1}{\includegraphics[width=\columnwidth]{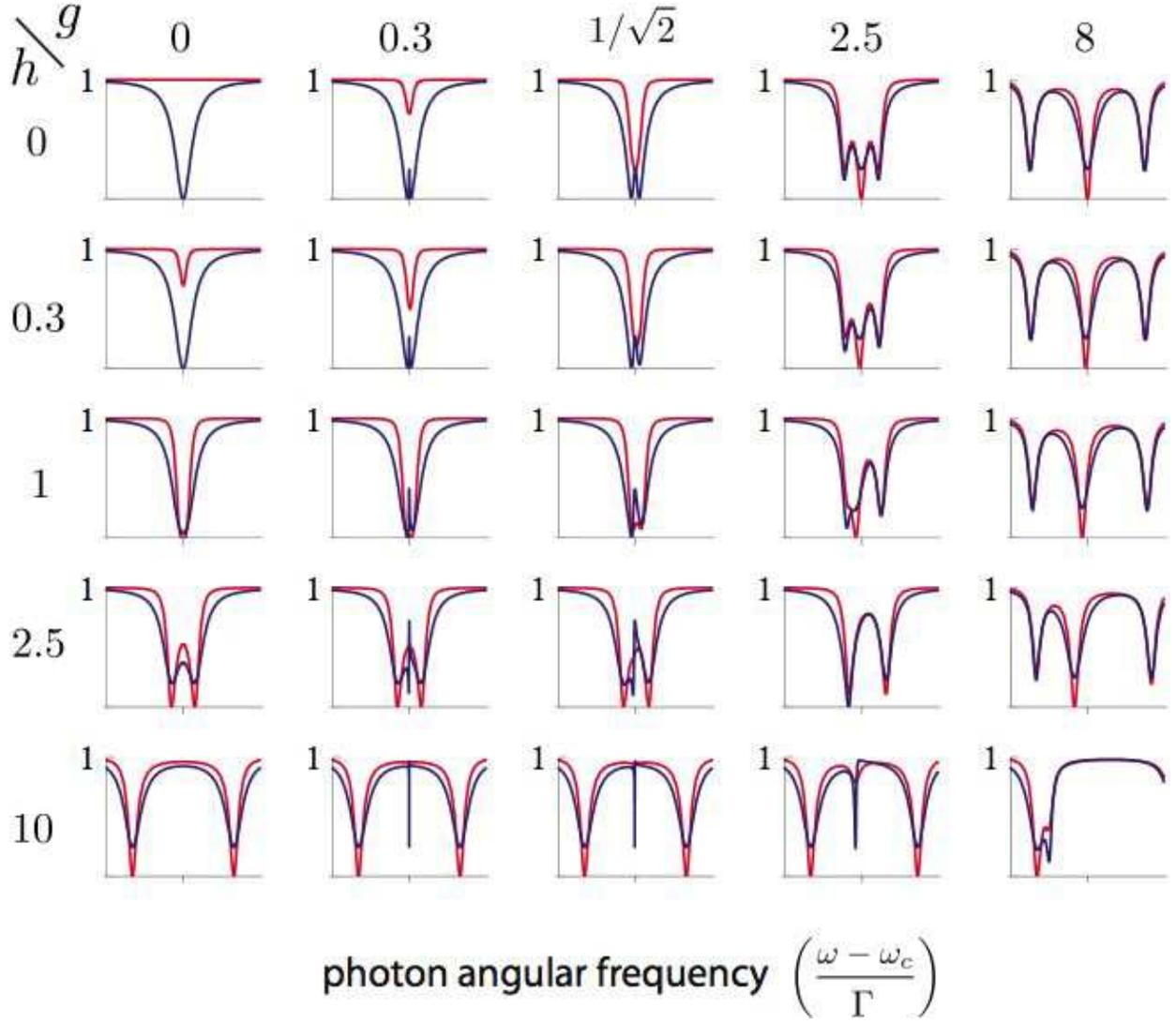}}
\caption{(Color online) Map of the transmission spectrum with intrinsic losses as $g$ and $h$ are varied. Red curves: only atom dissipation is non-zero ($1/\tau_q = \Gamma$ and $1/\tau_c=0$). Blue curves: only resonator dissipation is non-zero ($1/\tau_c = \Gamma$ and $1/\tau_q=0$). The intrinsic atom dissipation only affects resonances with atomic nature, and the intrinsic resonator dissipations only affects resonances with cavity nature. $h$ is real for this map. Also, $g_a = g_b$ and $g\equiv |g_a|$. The location of $\omega=\omega_c$ is indicated by the gray tick in each figure.}\label{Fi:TArrayDiss}
\end{figure}

This is clearly seen in all cases in Fig.~\ref{Fi:TArrayDiss}, where we introduce dissipations to all cases considered in Fig.~\ref{Fi:TArray}, which are lossless. We choose dissipations that are either purely from the atoms ($1/\tau_q = \Gamma$, $1/\tau_c=0$), or from the cavities ($1/\tau_c = \Gamma$, $1/\tau_q=0$).  
%As a concrete example, we consider the case of $h=10 \Gamma$ and $g=2.5 \Gamma$. The middle resonance is of purely atomic nature, and thus a small atom dissipation completely suppresses this resonance. This resonance is insensitive to a small cavity dissipation. The left resonance is of purely cavity nature, and thus is insensitive to the atom dissipation but is slightly suppressed by the cavity dissipation. The right resonance, being of largely cavity nature, is also insensitive of the atom dissipation.

\begin{figure}[htb]
%\scalebox{1.4}{\includegraphics{TRowg2p5h10Diss.eps}}
\scalebox{1}{\includegraphics[width=\columnwidth]{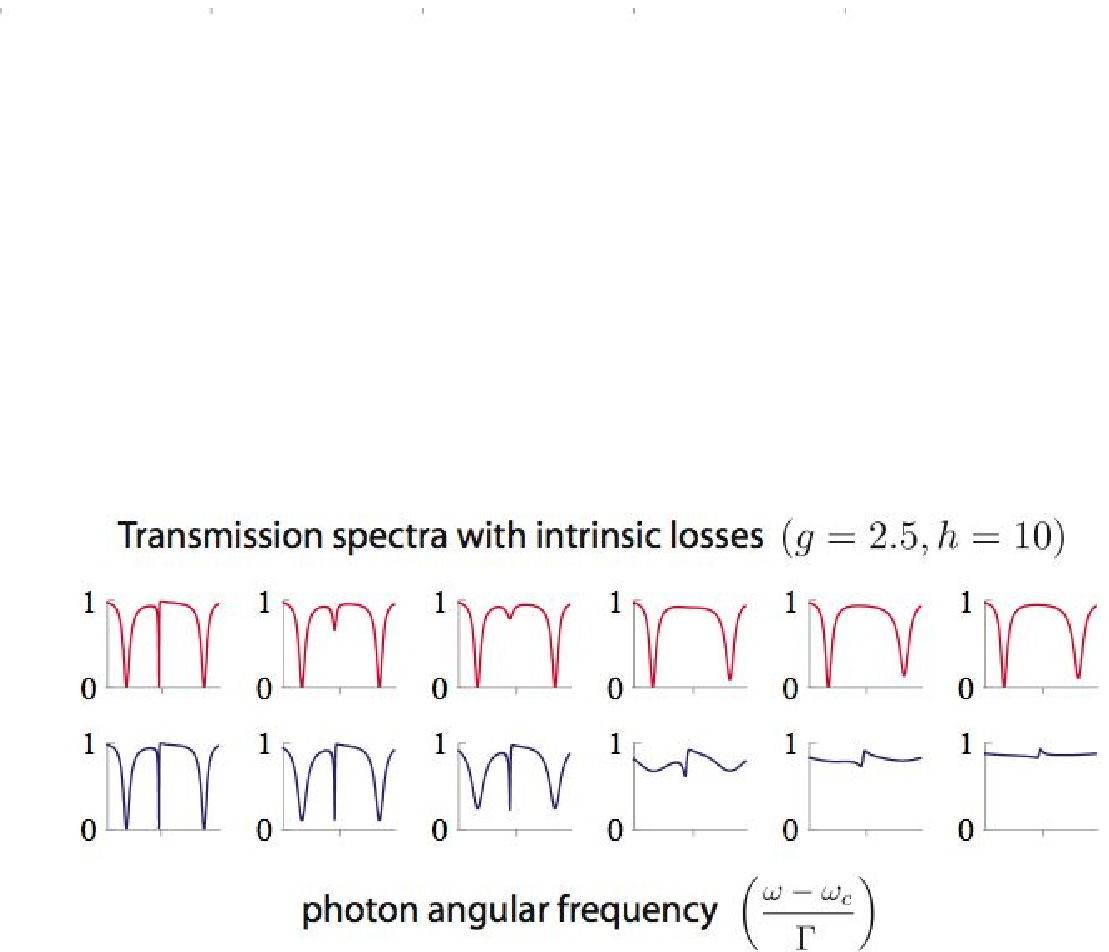}}
\caption{(Color online) The transmission spectrum for fixed $g$ and $h$ as the intrinsic atom and resonator losses are increased, respectively. From left to right: $0$, $0.5\Gamma$, $\Gamma$, $5\Gamma$, $10\Gamma$, and $20\Gamma$. Resonances of large cavity nature are less affected by the atom dissipation;  similarly, resonances of large atomic nature are less affected by the cavity dissipation. The location of $\omega=\omega_c$ is indicated by the gray tick in each figure.}\label{Fi:TRowDiss}
\end{figure}

%The nature of the resonances are valuable for understanding the transmission spectrum at the regime of large dissipations. 
As a more detailed example, Fig.~\ref{Fi:TRowDiss} plots the effects of increasing dissipations for the same case of $h=10 \Gamma$ and $g=2.5 \Gamma$. The  left resonance, which is of purely cavity nature, is essentially unaffected by even very large atom dissipation. The right resonance, which has a little atomic nature, has its transmission minimum gradually lifted and its width slightly broadened at large atom dissipation. The cavity dissipation, on the other hand, strongly affects all three resonances, since the phase matched WGM has weights at each resonance. Among the three resonances, the middle one is least affected, since it is primarily atomic. At large cavity dissipations, it still leaves a small signature in the transmission spectra while the contributions from the other two resonances are no longer visible. Thus, the nature of the resonance influences the properties of the transmission spectrum even in the regime of large dissipation. 
\section{The effects of complex $h$}\label{Sec:Complexh}

When $g_a=g_b$, all the amplitudes in Eq.~(\ref{E:th})-(\ref{E:eqh}) depend upon the phase of $h$ only but not of $g_a$. In the above discussion, for concreteness, we have assumed $h$ to be real. Here we consider the effects of a complex $h$ on the transport properties.

With $g_a$ and $g_b$ real, the spectral symmetry properties mentioned in Eq.~\eqref{E:General} in Sec.~\ref{Sec:GeneralProperties} now reads 
\begin{equation}\label{E:GeneralTSymmetry}
t^*\left(-\delta\omega, -\Delta, -\delta\theta\right) = t(\delta\omega, \Delta, \delta\theta),
\end{equation}where $\delta\theta\equiv \theta_h - \frac{\pi}{2}$. The spectral symmetry is determined by the phase of $h$ only.

%Since the phase is cyclic, one anticipates the spectra satisfy some symmetry property. Using the analytic expression of the transmission amplitude in Eq.~(\ref{E:th}), one could prove that the transmission amplitude satisfies the following symmetry condition, \emph{regardless of the magnitude of the cavity and atom dissipations}:
%The 
%\begin{equation}\label{E:GeneralTSymmetry}
%t^*\left(-\delta\omega, -\delta\theta, -\Delta\right) = t(\delta\omega, \delta\theta, \Delta),
%\end{equation}where $\delta\omega\equiv \omega-\omega_c$, $\delta\theta\equiv \theta_h - \frac{\pi}{2}$, and $\Delta\equiv\Omega-\omega_c$. This symmetry property states that the transmission spectrum for a system with parameters of $\delta\theta$ and $\Delta$ is mirror image to the transmission spectrum of another system with the parameters of $-\delta\theta$ and $-\Delta$, with respect to $\omega=\omega_c$. The same conclusion holds for other spectra in Eq.~(\ref{E:rh})-(\ref{E:eqh}). 

\begin{figure}[htb]
%\scalebox{1.4}{\includegraphics{TArrayg2p5h2p5Complex.eps}}
\scalebox{1}{\includegraphics[width=\columnwidth]{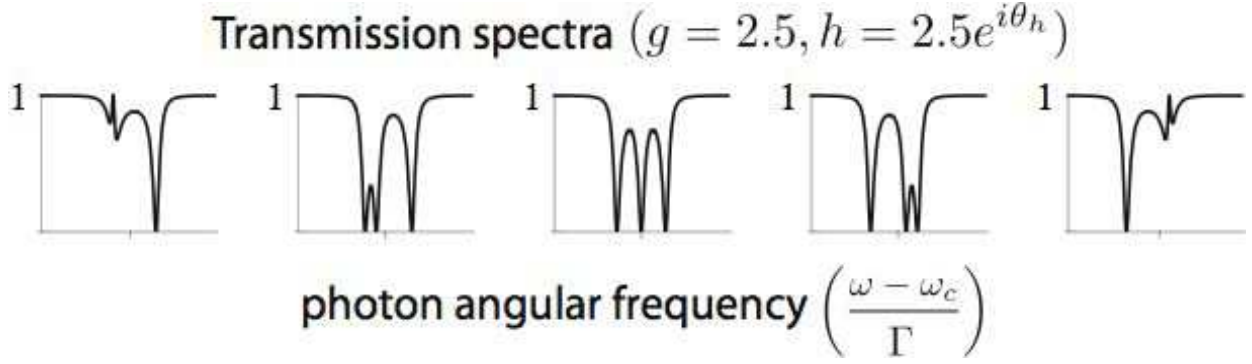}}
\caption{The transmission spectrum when $h$ is complex. $g=2.5\Gamma$. $h=2.5 e^{i\theta_h}\Gamma$. From left to right: $\theta=0$, $\pi/4$, $\pi/2$, $3\pi/4$, and $\pi$. The spectrum of $\theta=0$ is mirror image to that of $\theta=\pi$, and $\theta=\pi/4$ to $3\pi/4$. The spectrum of $\theta=\pi/2$ is symmetric with respect to $\omega=\omega_c$. The location of $\omega=\omega_c$ is indicated by the gray tick in each figure.}\label{Fi:TRowhComplex}
\end{figure}

Fig.~\ref{Fi:TRowhComplex} shows a series of transmission spectrum with $g=2.5 \Gamma$, and $h=2.5 e^{i \theta_h} \Gamma$ and $\Delta=0$, \emph{i.e.}, the cavity and the atom are in-tuned:
\begin{enumerate}
\item When $\theta_h=0$, two resonances of very different quality factor $Q$ are close to each other at $\omega < \omega_c$.
\item As $\theta_h=\pi/4$, the spectrum has three dips and is asymmetric with respect to $\omega=\omega_c$.
\item  When $\theta_h=\pi/2$, satisfying the condition that $\theta_h - 2\theta_g = \frac{\pi}{2} + n\pi$ with $n$ an integer, the spectrum is symmetric with respect to $\omega=\omega_c$.
\item The spectrum of $\theta_h=3\pi/4$ is the mirror image of the spectrum of $\theta_h=\pi/4$ with respect to $\omega=\omega_c$.
\item Finally, the spectrum of $\theta_h=\pi$ is the mirror image of that of $\theta_h=0$  with respect to $\omega=\omega_c$. 
\end{enumerate}

\begin{figure}[htb]
\scalebox{1}{\includegraphics{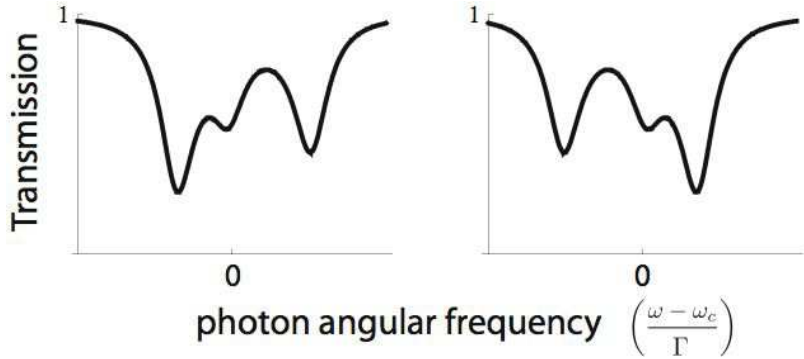}}
%\scalebox{1}{\includegraphics[width=\columnwidth]{symmetry.eps}}
\caption{The general symmetry property of the transmission spectrum when $\Delta\neq 0$. Left: $\Omega-\omega_c=2\Gamma$, $h=5 e^{i\pi/4}\Gamma$. Right: $\Omega-\omega_c=-2\Gamma$, $h=5 e^{i3\pi/4}\Gamma$. $1/\tau_q = 2\Gamma$, $1/\tau_c =\Gamma$, and $g=3\Gamma$ for both plots.}\label{Fi:symmetry}
\end{figure}

A general case is plotted in Fig.~\ref{Fi:symmetry}, with atom-cavity detuning $\Delta\neq 0$, and with finite atom and resonator dissipations. The transmission spectra with ($-\Delta, -\theta$) and with ($\Delta, \theta$) are mirror-imaged with respect to $\omega=\omega_c$.

%The left plot is $\Delta=\Omega-\omega_c=2\Gamma$, $h=5 e^{i\pi/4}\Gamma$ ($\delta\theta = -\pi/4$), while the right plot has  $\Delta=\Omega-\omega_c=-2\Gamma$, $h=5 e^{i3\pi/4}\Gamma$ ($\delta\theta = +\pi/4$). For both plots, $1/\tau_q = 2\Gamma$, $1/\tau_c =\Gamma$, and $g=3\Gamma$. These parameters satisfy Eq.~\eqref{E:GeneralTSymmetry}, and the spectra are mirror image to each other.

\section{Effects of Detuning ($\Omega\neq\omega_c$)}\label{Sec:Detuning}

\begin{figure}[htb]
%\scalebox{1.4}{\includegraphics{Detune.eps}}
\scalebox{1}{\includegraphics[width=\columnwidth]{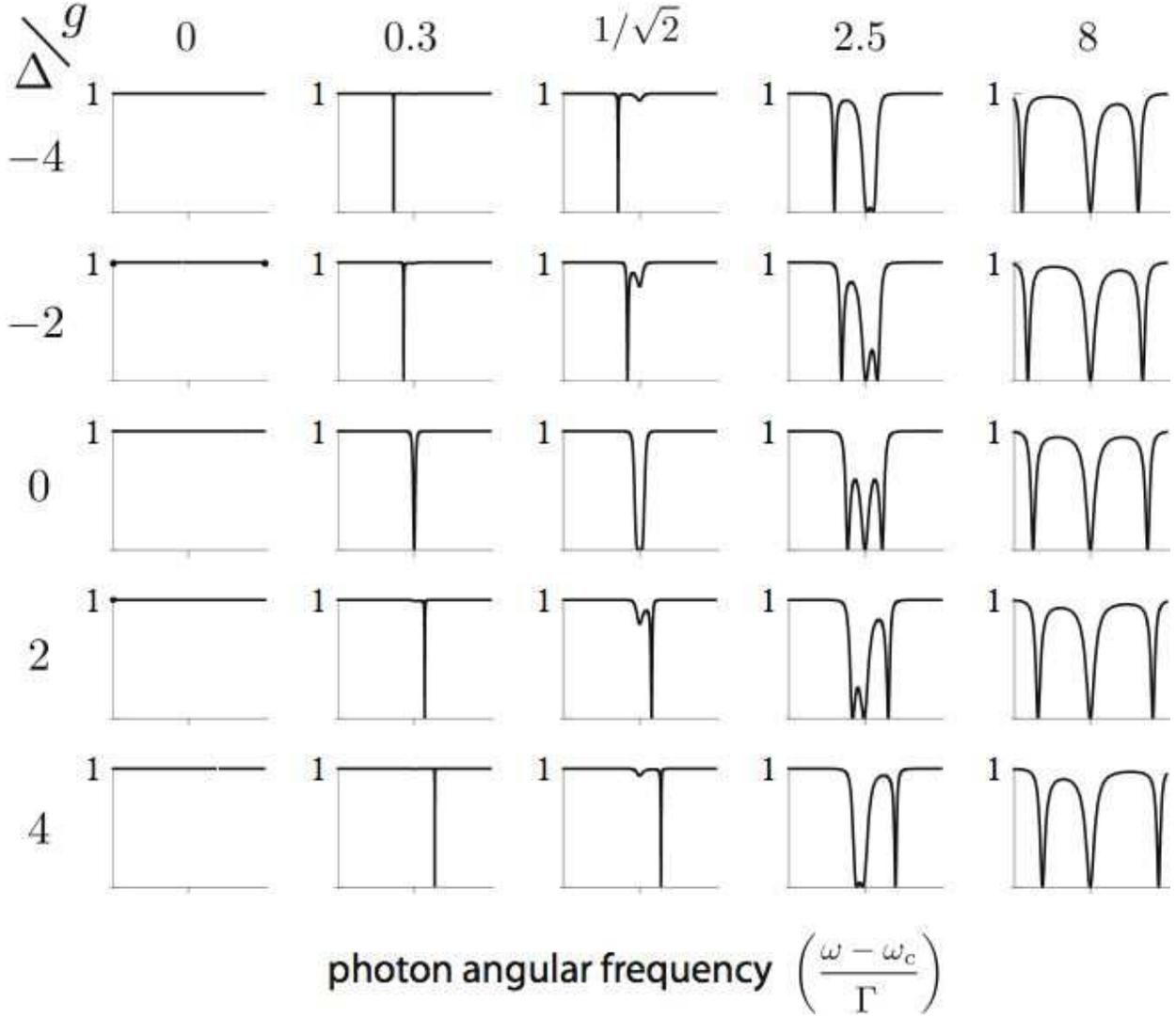}}
\caption{Transmission spectrum for the lossless detuned case. $h=0$, $g_a = g_b$ and $g\equiv |g_a|$. The location of $\omega=\omega_c$ is indicated by the gray tick in each figure.}\label{Fi:Detune}
\end{figure}

In this section, we discuss the effects of detuning between the resonator and the atom, with $h=0$. Fig.~\ref{Fi:Detune} plots the transmission spectrum for the lossless case as $g$ or $\Delta$ is varied. We note that the spectrum is always asymmetrical with respect to $\omega=\omega_c$ when both $\Delta$ and $g$ are non-zero.
%, which is a direct consequence of the spectral symmetry property of Eq.~\eqref{E:th}. 
We now discuss these spectra plots, as organized in a matrix form.

\subsection{The first row: large detuning with $|\Delta|\gg\Gamma$}
\begin{enumerate}
\item When $g \ll \Delta$ ($g=0.3\Gamma$ as plotted), the atom is decoupled from the system and creates a narrow atomic resonance at $\omega=\Omega$. The background is the transmission of waveguide-ring resonator subsystem that has unity transmission.
\item When $g$ is increased ($g=1/\sqrt{2}\Gamma$ as plotted), 
%but with $2\sqrt{2}g \leq \Delta$, where $2\sqrt{2}g$ is the Rabi frequency, and with $\sqrt{2} g < \Gamma$, 
the atom becomes weakly coupled to the waveguide-ring resonator subsystem. Analytically, such a weak coupling regime occurs in the parameter range $2\sqrt{2}g \lesssim\Delta$, where $2\sqrt{2}g$ is the Rabi frequency, and $\sqrt{2} g \lesssim \Gamma$. In this weak coupling regime, the transmission spectrum exhibits an atomic resonance dip at $\omega = \Omega$. Also, the weak coupling of the atom and the cavity results in a small scattering between the two WGM, and consequently a small dip is present at the resonant frequency $\omega=\omega_{c}$. Notice that the transmission dip at the cavity resonant frequency does not reaches zero at its minimum. 
\item For large $g$ such that $2\sqrt{2}g \gtrsim \Delta$ and $\sqrt{2}g \gtrsim \Gamma$, the atom is strongly coupled to the waveguide-resonator subsystem, and the transmission spectrum develops into three resonance dips, each reaching zero at its minimum. . 
\end{enumerate}

\subsection{Second column: decoupled and weakly coupled atom}
%\subsection{Second column: weakly coupled atom ($2\sqrt{2}g \lesssim \Delta$ and $\sqrt{2} g \lesssim \Gamma$)}

For decoupled ($g \ll \Delta$) and weakly coupled ($2\sqrt{2}g \lesssim\Delta$, $\sqrt{2} g \lesssim \Gamma$) atom, the atom creates a narrow atomic resonance at $\omega=\Omega$, and the location of the resonance moves with $\Delta$. At $\Delta =0$, the weakly coupled atom mixes with the cavity resonance so the line width is slightly increased, compared with that of the $\Delta\neq 0$ cases.

\begin{figure}[htb]
%\scalebox{1.4}{\includegraphics{DetuneLoss.eps}}
\scalebox{1}{\includegraphics[width=\columnwidth]{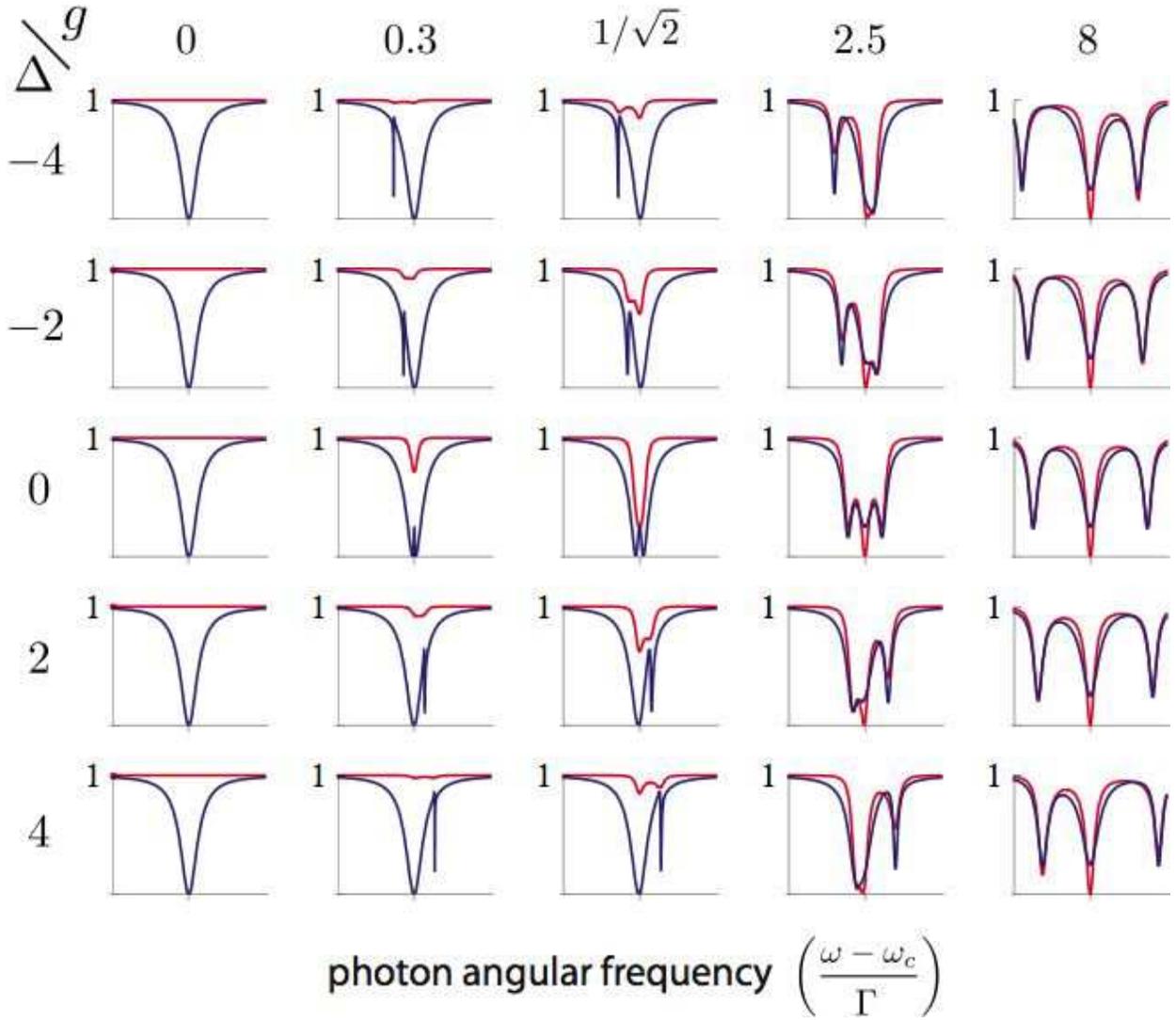}}
\caption{(Color online) Transmission spectrum for the detuned case with losses. Red curves: only atom dissipation is non-zero ($1/\tau_q = \Gamma$ and $1/\tau_c=0$). Blue curves: only resonator dissipation is non-zero ($1/\tau_c = \Gamma$ and $1/\tau_q=0$). $h=0$, $g_a = g_b$ and $g\equiv |g_a|$. The location of $\omega=\omega_c$ is indicated by the gray tick in each figure.}\label{Fi:DetuneLoss}
\end{figure}

For intermediate values of $g$ and $\Delta$, the exact lineshape and the weight of the nature of a resonance has to be computed using the exact expressions. Fig.~\ref{Fi:DetuneLoss} plots the effects of intrinsic losses of the resonator and the atom to the transmission spectrum. Again, the resonator loss strongly suppresses a resonance of cavity nature, and the atomic loss strongly suppresses a resonance of atom nature.

\section{Summary}\label{Sec:Summary}
We have provided a full quantum mechanical approach to treat the coupled waveguide-ring resonator-atom system, and derived the analytic solutions for the single-photon transport. The real-space approach outlined in this article can be generalized straightforwardly to treat cases such as  cascaded multi-ring resonator, multi-atom, or multi-port configuration that is relevant to applications of add-drop filter, single-photon switching and delay lines. Our formalism can also provide a starting point for treatment of pulse propagation in these more complicated systems.

\begin{acknowledgments}
J.-T. Shen acknowledges informative discussions with K. Srinivasan at NIST, and S. Chiow at Stanford. S. Fan acknowledges financial support by the David and Lucile Packard Foundation.
\end{acknowledgments}

\appendix
\section{Symmetry transformations of the Hamiltonian}\label{A:SymmetryTransformations}

The relations between the coupling constants are not arbitrary but rather are constrained by the symmetry of the Hamiltonian. The symmetries of interest for our configurations are mirror symmetry and the time-reversal symmetry. In this section, we consider these symmetries that constrain the form of the Hamiltonian, and establish the relations between the coupling constants. Note that the specific form of the relations depend upon the choice of the representation of the fields.

\subsection{Waveguide mode operators}
\subsubsection{Mirror symmetry}
We start with the waveguide mode operators. The following general considerations apply to any single-mode waveguide that obeys mirror symmetry. For such a waveguide, one can always choose a mirror plane perpendicular to the waveguide such that the dielectric function $\epsilon(x, y,z)$ satisfies
\begin{equation}
\epsilon(-x, y, z) = \epsilon(x, y,z),
\end{equation}where the $x$-axis is along the direction of the waveguide. From the Maxwell's equations
\begin{align}\label{E:Maxwell}
\nabla \times \vec{E}(x, y,z, t) &= -\mu \frac{\partial}{\partial t}\vec{H}(x, y,z, t),\notag\\
\nabla \times \vec{H}(x, y,z, t) &= +\epsilon(x, y,z) \frac{\partial}{\partial t}\vec{E}(x, y,z, t),
\end{align}it is straightforward to show that if the fields of the following form
\begin{align}\label{E:RightMode}
\vec{E}_R(x, y,z, t) &\equiv \vec{A}e^{i k_x x - i\omega t}\equiv (A_x(y, z), A_y(y,z), A_z(y, z)) e^{i k_x x - i\omega t}\notag\\
&\equiv \vec{E}_R(x, y,z) e^{- i\omega t},\notag\\
\vec{H}_R(x, y,z, t) &\equiv \vec{B}e^{i k_x x - i\omega t}\equiv (B_x(y, z), B_y(y,z), B_z(y, z)) e^{i k_x x - i\omega t}\notag\\
&\equiv \vec{H}_R(x, y,z) e^{- i\omega t}
\end{align}is a solution of the Maxwell's equations for the waveguide, then the following fields are also a solution:
\begin{align}\label{E:LeftMode}
\vec{E}_L(x, y,z, t) &\equiv (-A_x(y, z), A_y(y,z), A_z(y, z)) e^{-i k_x x - i\omega t}\notag\\
&\equiv \vec{E}_L(x, y,z)e^{- i\omega t},\notag\\
\vec{H}_L(x, y,z, t) &\equiv (B_x(y, z), -B_y(y,z), -B_z(y, z)) e^{-i k_x x - i\omega t}\notag\\
&\equiv \vec{H}_L(x, y,z)e^{- i\omega t}.
\end{align}

The two sets of eigenmodes defined by Eqs.~\eqref{E:RightMode} and \eqref{E:LeftMode} obey the mirror-symmetry transformation: 
\begin{align}
\pi_x\left[\vec{E}_R(\pi_x \vec{r})\right] = \vec{E}_L(\vec{r}),\notag\\
\pi_x\left[\vec{H}_R(\pi_x \vec{r})\right] = \vec{H}_L(\vec{r}),
\end{align}where $\vec{r}\equiv (x, y, z)$, \emph{i.e.},
%and the mirror-transformation operator $\pi_{x}$ is defined such that:
\begin{align}\label{E:Rules}
\pi_x (x, y, z) &\equiv  (-x, y, z),\notag\\
\pi_x (E_x, E_y, E_z) &\equiv (-E_x, E_y, E_z),\notag\\
\pi_x (H_x, H_y, H_z) &\equiv (H_x, -H_y, -H_z)
\end{align}

In the second-quantization form, one thus defines a mirror-transormation operator $\pi_{x}$ such that~\cite{Greiner:1996}
\begin{equation}
\pi_x c_R^{\dagger}(x)\pi_x = c_L^{\dagger}(-x).
\end{equation}Note that the mirror-transformation operator $\pi_x$ is both unitary and  Hermitian
\begin{equation}
\pi_x^{-1} = \pi_{x}^{\dagger}=\pi_x,
\end{equation}due to the choice of the relative phase of $\vec{E}_L$ with respect to $\vec{E}_R$ in Eq.~\eqref{E:LeftMode}~\cite{Sakurai:1994}. Also, in the second quantization form, $\pi_x$ operates only on operators, not on numbers or functions.

As an example, one can show that the mirror transformation of a right-moving photon is a left-moving photon:
\begin{align}
&\pi_x \int dx f(x) c_R^{\dagger}(x)|\emptyset\rangle\notag\\
=& \int dx f(x) \left[\pi_x c_R^{\dagger}(x) \pi_x^{-1}\right] \pi_x|\emptyset\rangle\notag\\
=&  \int dx f(x) c_L^{\dagger}(-x)|\emptyset\rangle\notag\\
=& \int dx f(-x) c_L^{\dagger}(x)|\emptyset\rangle,
\end{align}where, without losing generality, the ground state is assumed to have positive parity: $\pi_x|\emptyset\rangle= |\emptyset\rangle$.
\subsubsection{Time-reversal symmetry}

From the Maxwell's equations Eqs.~\eqref{E:Maxwell}, one can show that if $\eta\vec{E}_R(\vec{r})$ and $\eta\vec{H}_R(\vec{r})$ is an eigenmode, where $\vec{E}_R$ and $\vec{H}_R$ are the fields in Eq.~\eqref{E:RightMode} and $\eta$ is an arbitrary complex number, then
\begin{align}\label{E:RightTimeReversal}
\eta^*\vec{E}_{R}^*(\vec{r})&=\eta^*\vec{A}^* e^{-i k_x x}\notag\\
-\eta^*\vec{H}_{R}^*(\vec{r}) &= \eta^*\left(-\vec{B}^*\right) e^{-i k_x x}
\end{align}
is also an eigenmode~\cite{Haus:1984}. $\vec{E}_{R}^*(\vec{r})$ and $-\vec{H}_{R}^*(\vec{r})$ are the time-reversed fields corresponding to $\vec{E}_R(\vec{r})$ and $\vec{H}_R(\vec{r})$. Since the waveguide is single-moded, we must have
\begin{align}
\vec{E}_R^*(\vec{r}) &= \vec{E}_L(\vec{r}),\notag\\
-\vec{H}_R^*(\vec{r}) &= \vec{H}_L(\vec{r}),
\end{align}where we have chosen the proportionality constant to be $1$.

Thus we define a time-reversal operator $T$
\begin{align}
T\left[\eta \vec{E}_R\right] &\equiv \eta^* \vec{E}_L,\notag\\
T\left[\eta \vec{H}_R\right] &\equiv \eta^* \vec{H}_L.
\end{align}In the second quantization form, one defines a time-reversal operator $T$ such that
\begin{equation}
T\left(\eta c_R^{\dagger}(x)\right)T^{-1} = \eta^* c_L^{\dagger}(x).
\end{equation}
As an example, for any right-moving one-photon state, the time-reversed state is
\begin{align}
&T \int dx f(x) c_R^{\dagger}(x)|\emptyset\rangle\notag\\
=& \int dx f^*(x) \left[T c_R^{\dagger}(x) T^{-1}\right] T|\emptyset\rangle\notag\\
=&  \int dx f^*(x) c_L^{\dagger}(x)|\emptyset\rangle,
\end{align}where, without losing generality, we have assumed $T|\emptyset\rangle= |\emptyset\rangle$.

As a side remark, we note that combining the mirror and time-reversal symmetries allow us to make further statement on the fields. For example, $A_x$ can be chosen to be purely imaginary, while $A_y$ and $A_z$ can be chosen to be purely real.

\subsection{Single-mode cavity}
For a single-mode cavity that obeys mirror symmetry, one must have~\cite{Fan:1999}
\begin{align}
\pi_x\left[\vec{E}(\pi_x \vec{r})\right] = \pm\vec{E}(\vec{r}),\notag\\
\pi_x\left[\vec{H}(\pi_x \vec{r})\right] = \mp\vec{H}(\vec{r}),
\end{align}where the fields that transform with the upper(lower) sign is called an even(odd) mode. Thus, if $a$ is the creation operator of the single-mode, one has
\begin{equation}
\pi_x a^{\dagger}\pi_x = \pm a^{\dagger},
\end{equation}where $+$($-$) sign is for even(odd) mode. Similarly, since a single mode must be mapped into itself, one has
\begin{equation}
T a^{\dagger}T^{-1} = a^{\dagger}.
\end{equation}

As an example, we now show that the symmetry considerations allow us to use a single real number to characterize the coupling between a single-mode cavity and the waveguide. The interaction between them can be written as
\begin{equation}\label{E:SingleModeInteraction}
H_{I} \equiv \int dx \delta(x) \left[V_1 c_R^{\dagger}(x) a + V_1^* a^{\dagger} c_R(x)\right] + \int dx \delta(x) \left[V_2 c_L^{\dagger}(x) a + V_2^* a^{\dagger} c_L(x)\right].
\end{equation}By writing down the interaction, we only require it to be Hermitian.

The mirror-symmetry invariance requires $\pi_x H_{I}\pi_x = H_I$:
\begin{align}
&\pi_x H_I \pi_x\notag\\
=& \int dx \delta(x) \left[V_1 \pi_xc_R^{\dagger}(x)\pi_x\pi_x a \pi_x+ V_1^* \pi_x a^{\dagger}\pi_x \pi_x c_R(x)\pi_x\right]\notag\\ 
&+ \int dx \delta(x) \left[V_2 \pi_x c_L^{\dagger}(x)\pi_x \pi_x a \pi_x + V_2^* \pi_x a^{\dagger}\pi_x \pi_xc_L(x)\pi_x\right]\notag\\
=& \int dx \delta(x)\left(\pm\right) \left[V_1 c_L^{\dagger}(-x) a + V_1^* a^{\dagger} c_L(-x)\right] + \int dx \delta(x) \left(\pm\right)\left[V_2 c_R^{\dagger}(-x) a + V_2^* a^{\dagger} c_R(-x)\right]\notag\\
=&\int dx \delta(x) \left(\pm\right)\left[V_1 c_L^{\dagger}(x) a + V_1^* a^{\dagger} c_L(x)\right] + \int dx \delta(x) \left(\pm\right)\left[V_2 c_R^{\dagger}(x) a + V_2^* a^{\dagger} c_R(x)\right].
\end{align}Comparing with Eq.~\eqref{E:SingleModeInteraction}, one thus has $\pm V_1 = V_2$.

The time-reversal invariance requires $T H_I T^{-1} = H_I$:
\begin{align}
&T H_I T^{-1}\notag\\
=& \int dx \delta(x) \left[V_1^* T c_R^{\dagger}(x)T^{-1} T a T^{-1}+ V_1 T a^{\dagger}T^{-1} T c_R(x)T^{-1}\right]\notag\\ 
&+ \int dx \delta(x) \left[V_2^* T c_L^{\dagger}(x)T^{-1} T a T^{-1} + V_2 T a^{\dagger}T^{-1} T c_L(x)T^{-1}\right]\notag\\
=& \int dx \delta(x) \left[V_1^* c_L^{\dagger}(x) a + V_1 a^{\dagger} c_L(x)\right] + \int dx \delta(x) \left[V_2^* c_R^{\dagger}(x) a + V_2 a^{\dagger} c_R(x)\right].
\end{align}Comparing with Eq.~\eqref{E:SingleModeInteraction}, one concludes that $V_1^*=V_2$.

Combine the above results, one reaches the result that $V_1=V_2=V_1^*\equiv V$ for even mode, while $V_1=-V_2=-V_1^*\equiv iV$ for odd mode, where $V$ is a real number. Thus the coupling between a single-mode cavity and the waveguide could be represented by a single real number $V$, when the structure obeys mirror and time-reversal symmetry.

\subsection{WGM-type resonator}

We now consider a resonator which possesses azimuthal rotational symmetry and supports a pair of degenerate counter-propagating WGMs. Many resonators are of this type: two-dimensional dielectric cylinder, microsphere,  microdisk, microtoroid, and ring resonator. 

\subsubsection{Mirror symmetry}
We consider a two-dimensional cylindrically symmetric system. In the cylindrical coordinates $(\rho, \phi, z)$, if the fields of the following form
\begin{align}\label{E:CounterClockMode}
\vec{E}_m(\rho, \phi', z, t) &\equiv \vec{A}e^{i m \phi' - i\omega t}\equiv (A_\rho(\rho, z), A_\phi(\rho, z), A_z(\rho, z)) e^{i m\phi' - i\omega t}\notag\\
&\equiv \vec{E}_m(\rho, \phi', z) e^{- i\omega t},\notag\\
\vec{H}_m(\rho, \phi', z, t) &\equiv \vec{B}e^{i m \phi' - i\omega t}\equiv (B_\rho(\rho, z), B_\phi(\rho, z), B_z(\rho, z)) e^{i m\phi' - i\omega t}\notag\\
&\equiv \vec{H}_m(\rho, \phi', z) e^{- i\omega t}
\end{align}are a set of solutions of Maxwell's equations for the resonator, then the following fields are also a set of solutions :
\begin{align}\label{E:ClockMode}
\vec{E}_{-m}(\rho, \phi', z, t) &\equiv (A_\rho(\rho, z), -A_\phi(\rho, z), A_z(\rho, z)) e^{-i m\phi' - i\omega t}\notag\\
&\equiv \vec{E}_{-m}(\rho, \phi', z) e^{- i\omega t},\notag\\
\vec{H}_{-m}(\rho, \phi', z, t) &\equiv  (-B_\rho(\rho, z), B_\phi(\rho, z), -B_z(\rho, z)) e^{-i m\phi' - i\omega t}\notag\\
&\equiv \vec{H}_{-m}(\rho, \phi', z) e^{- i\omega t}.
\end{align}

The unit vectors in the cylindrical coordinates could be written in the Cartesian coordinates as
\begin{align}\label{E:bases}
\hat{\rho} &= \cos\phi' \,\hat{x}' + \sin\phi' \,\hat{y}',\notag\\
\hat{\phi} &= -\sin\phi' \,\hat{x}' + \cos\phi' \,\hat{y}', \notag\\
\hat{z} &=\hat{z},
\end{align}where the $x'$-axis is the azimuthal origin of $\phi'=0$, and needs not to be coincident with $x$-axis that is along the direction of the waveguide.

Thus, the two sets of eigenmodes of Eq.~\eqref{E:CounterClockMode} and Eq.~\eqref{E:ClockMode} in the Cartesian coordinates could be expressed as
\begin{align}
\vec{E}_m(\vec{r}) &= \left[\left(A_\rho \cos\phi' - A_\phi \sin\phi'\right)\,\hat{x}' + \left(A_\rho \sin\phi' + A_\phi \cos\phi'\right)\, \hat{y}' + A_z \hat{z}\right] e^{i m\phi'},\notag\\
\vec{H}_m(\vec{r}) &= \left[\left(B_\rho \cos\phi' - B_\phi \sin\phi'\right)\,\hat{x}' + \left(B_\rho \sin\phi' + B_\phi \cos\phi'\right)\, \hat{y}' + B_z \hat{z}\right] e^{i m\phi'},
\end{align}and 
\begin{align}
\vec{E}_{-m}(\vec{r}) &= \left[\left(A_\rho \cos\phi' + A_\phi \sin\phi'\right)\,\hat{x}' + \left(A_\rho \sin\phi' - A_\phi \cos\phi'\right)\, \hat{y}' + A_z \hat{z}\right] e^{-i m\phi'},\notag\\
\vec{H}_{-m}(\vec{r}) &= \left[\left(-B_\rho \cos\phi' - B_\phi \sin\phi'\right)\,\hat{x}' + \left(-B_\rho \sin\phi' + B_\phi \cos\phi'\right)\, \hat{y}' - B_z \hat{z}\right] e^{-i m\phi'},
\end{align}respectively.

An azimuthal rotational symmetric structure also possesses mirror symmetry. The choice of mirror plane relative to the $x'$-axis however affects the specific form of transformation between the fields. Here we examine this degree of freedom. 

As before, let $x$-axis be along the waveguide direction, and the mirror plane is the $y$-$z$ plane that perpendicular to the $x$-axis. If the angle between the $x'$-axis and the $x$-axis is $\phi_0$ (see Fig.~\ref{Fi:PhiDefinition}), the bases of Eqs.~\eqref{E:bases} in $x-y$ coordinate become
\begin{align}\label{E:bases2}
\hat{\rho} &=\cos(\phi'+\phi_0)\,\hat{x} + \sin(\phi'+\phi_0)\,\hat{y},\notag\\
\hat{\phi} &= -\sin(\phi'+\phi_0) \,\hat{x} + \cos(\phi'+\phi_0) \,\hat{y}, \notag\\
\hat{z} &=\hat{z},
\end{align}and the fields become
\begin{align}
\vec{E}_m &= \left[\left(A_\rho \cos\phi - A_\phi \sin\phi\right)\,\hat{x} + \left(A_\rho \sin\phi + A_\phi \cos\phi\right)\, \hat{y} + A_z \hat{z}\right] e^{i m\phi}e^{-i m\phi_0},\notag\\
\vec{H}_m &= \left[\left(B_\rho \cos\phi - B_\phi \sin\phi\right)\,\hat{x} + \left(B_\rho \sin\phi + B_\phi \cos\phi\right)\, \hat{y}+ B_z \hat{z}\right] e^{i m\phi}e^{-i m\phi_0},\notag\\
\vec{E}_{-m} &= \left[\left(A_\rho \cos\phi + A_\phi \sin\phi\right)\,\hat{x} + \left(A_\rho \sin\phi - A_\phi \cos\phi\right)\, \hat{y} + A_z \hat{z}\right] e^{-i m\phi}e^{+i m\phi_0},\notag\\
\vec{H}_{-m} &= \left[\left(-B_\rho \cos\phi - B_\phi \sin\phi\right)\,\hat{x}+ \left(-B_\rho \sin\phi + B_\phi \cos\phi\right)\, \hat{y} - B_z \hat{z}\right] e^{-i m\phi}e^{+i m\phi_0},
\end{align}where $\phi\equiv\phi'+\phi_0$ is the azimuthal angle of the point $\vec{r}$ in the $x$-$y$ coordinate. The mirror transformation that maps $x\rightarrow -x$ is equivalent to the mapping $\phi\rightarrow \pi-\phi$, while with $\rho$ and $z$ unchanged. The fields thus transform according to
\begin{align}
\pi_{x} \left[\vec{E}_{m}\left(\pi_x\vec{r}\right)\right] &= (-1)^m e^{-2 i m \phi_0}\vec{E}_{-m}\left(\vec{r}\right),\notag\\
\pi_{x} \left[\vec{H}_{m}\left(\pi_x\vec{r}\right)\right] &= (-1)^m e^{-2 i m \phi_0}\vec{H}_{-m}\left(\vec{r}\right),
\end{align}where the transformation rules of $\pi_x$ on fields are given by Eq.~\eqref{E:Rules}.

In the second quantization form, accordingly one has
\begin{align}
\pi_x a^{\dagger} \pi_x &= (-1)^m e^{-2 i m \phi_0} b^{\dagger}=e^{i 2 m \bar{\phi}} b^{\dagger},\notag\\
\pi_x b^{\dagger} \pi_x &= (-1)^m e^{+2 i m \phi_0} a^{\dagger}=e^{-i 2 m \bar{\phi}}a^{\dagger},
\end{align}where $\bar{\phi}\equiv \pi/2 -\phi_0$ is the angle between the $x'$-axis and the mirror plane ($y$-$z$ plane). In the special case where the $x'$-axis and the $x$-axis are coincident ($\phi_0=0$), one has
\begin{align}
\pi_x a^{\dagger} \pi_x &= (-1)^m b^{\dagger},\notag\\
\pi_x b^{\dagger} \pi_x &= (-1)^m a^{\dagger};
\end{align}where in the special case where the $x'$-axis and the $y$-axis are coincident ($\phi_0=\pi/2$), one has
\begin{align}
\pi_x a^{\dagger} \pi_x &=  b^{\dagger},\notag\\
\pi_x b^{\dagger} \pi_x &=  a^{\dagger}.
\end{align}

\subsubsection{Time-reversal symmetry}
The discussion of the transformation by the time-reversal operator $T$ is the similar to that of the waveguide case. If the fields $\eta\vec{E}_{m}(\vec{r})$ and $\eta\vec{H}_{m}(\vec{r})$ are a set of solutions, then $\eta^*\vec{E}^*_{m}(\vec{r})$ and $-\eta^*\vec{H}^*_{m}(\vec{r})$ are also a set of solutions. From the functional form, one concludes that
\begin{align}
\vec{E}^*_{m}(\vec{r}) &=\vec{E}_{-m}(\vec{r}),\notag\\
-\vec{H}^*_{m}(\vec{r}) &=\vec{H}_{-m}(\vec{r}),
\end{align}where we have chosen the proportionality constant to be 1. We thus define a time-reversal operator $T$ such that
\begin{align}
T\left[\eta\vec{E}_{m}\right] &= \eta^*\vec{E}_{-m},\notag\\
T\left[\eta\vec{H}_{m}\right] &= \eta^*\vec{H}_{-m}.
\end{align}

In the second quantization form, one defines a time-reversal operator such that
\begin{equation}
T\left(\eta a^{\dagger}\right)T^{-1} = \eta^* b^{\dagger}.
\end{equation}

As a side remark, we note that combining the mirror and time-reversal symmetries allow us to make further statement on the fields. For example, $A_\phi$  can be chosen to be purely imaginary, while $A_\rho$, and $A_z$ can be chosen to be purely real.

\subsubsection{Waveguide-ring resonator interactions}

The general form of the waveguide-ring resonator interaction can be written as
\begin{equation}
H_I = \int dx \, \delta(x)[V_a c_R^{\dagger}(x)a + V_a^* a^{\dagger} c_R(x)] + \int dx \, \delta(x)[V_b c_L^{\dagger}(x)b + V_b^* b^{\dagger} c_L(x)],
\end{equation}where we only require it to be Hermitian. The symmetry invariance further impose constraints on the coupling constants:\\
The mirror-symmetry invariance requires $\pi_x H_{I}\pi_x = H_I$:
\begin{align}
&\pi_x H_I \pi_x\notag\\
=& \int dx \, \delta(x)[V_a c_L^{\dagger}(-x) e^{-i 2 m \bar{\phi}} b + V_a^* e^{+i 2 m \bar{\phi}}b^{\dagger} c_L(-x)]\notag\\
& + \int dx \, \delta(x)[V_b c_R^{\dagger}(-x) e^{+i 2 m \bar{\phi}} a + V_b^* e^{-i 2 m \bar{\phi}} a^{\dagger} c_R(-x)],
\end{align}which yields $V_b e^{i 2 m \bar{\phi}} = V_a$, where $\bar{\phi}=\pi/2 -\phi_0$, as previously defined.\\
The time-reversal symmetry invariance requires $T H_I T^{-1}=H_I$:
\begin{align}
&T H_I T^{-1}\notag\\
=&  \int dx \, \delta(x)[V_a^* c_L^{\dagger}(x)b + V_a b^{\dagger} c_L(x)] + \int dx \, \delta(x)[V_b^* c_R^{\dagger}(x)a + V_b a^{\dagger} c_R(x)],
\end{align}which yields $V_a = V_b^*$.

Combine the results together, one has
\begin{align}
V_a &= V e^{i m \bar{\phi}},\notag\\
V_b &= V e^{- i m \bar{\phi}},
\end{align}where $V = |V_a|=|V_b|$ is a real number.

%As a side remark, we note that combining the mirror and time-reversal symmetries allow us to make further statement on the fields. For example, $A_\phi$  can be chosen to be purely imaginary, while $A_\rho$, and $A_z$ can be chosen to be purely real.

\section{Resonator-atom interactions}\label{A:RAInteraction}

In this section we outline the procedures to obtain the interaction between the resonator and the atom in the second quantized form. This derivation gives the general form of the interaction.

The interaction between the resonator and the atom is given by
\begin{equation}
H_{I} = - \mathbf{d}\cdot\mathbf{E},
\end{equation}where $\mathbf{d}= e \mathbf{r}$ is the electric dipole moment of the atom ($e>0$ is the magnitude of charge), and $\mathbf{E}$ is the electric field of the resonator modes. This expression is valid when the wavelength of interest is much longer than the dimension of the atom. The field is evaluated at the location of the dipole $\mathbf{r}=\mathbf{r}_0$. 

The atomic electric dipole operator could be expanded using the bases of the atom:
\begin{align}
\mathbf{d} &= \left(|-\rangle\langle -| + |+\rangle\langle +|\right) \mathbf{d}\left(|-\rangle\langle -| + |+\rangle\langle +|\right)\notag\\
&= \left( |+\rangle\langle +|\mathbf{d}|-\rangle\langle -|\right)+ \left(|-\rangle\langle -|\mathbf{d}|+\rangle\langle +|\right)\notag\\
&\equiv \vec{d} |+\rangle\langle-| + \vec{d}^* |-\rangle\langle+|\notag\\
&=  \vec{d}\sigma_{+} + \vec{d}^* \sigma_{-},
\end{align} with $\vec{d} = e\langle +|\mathbf{r}|-\rangle$. The terms $\langle +|\mathbf{d}|+\rangle=\langle -|\mathbf{d}|-\rangle$ vanish since $\mathbf{d}$ has odd parity. Here the atomic basis $|-\rangle$ and $|+\rangle$ are taken as eigenstates of parity. Thus the atomic electric dipole operator is entirely off-diagonal in the atomic bases. 

The electric field operator is expanded using the eigenmodes of the resonator. We will treat two cases here: the single-mode cavity and a ring-resonator.

For a single-mode cavity, the electric field operator is
\begin{equation}
\mathbf{E}(\mathbf{r}) = a \vec{\phi}(\mathbf{r}) +  a^{\dagger} \vec{\phi}^*(\mathbf{r}), 
\end{equation}where $\vec{\phi}$ is the field profile of the mode, up to a global phase that does not affect the physics. Consequently, the interaction takes the following form:
\begin{align}
H_I &= -\left(a \vec{\phi}(\mathbf{r}) +  a^{\dagger} \vec{\phi}^*(\mathbf{r})\right)\cdot\left(\vec{d}\sigma_{+} + \vec{d}^* \sigma_{-}\right)\notag\\
&= -\left(\vec{\phi}(\mathbf{r})\cdot\vec{d}\right)  a \sigma_{+}  -\left(\vec{\phi}^*(\mathbf{r})\cdot\vec{d}^*\right)  a^{\dagger} \sigma_{-} -\left(\vec{\phi}(\mathbf{r})\cdot\vec{d}^*\right) a \sigma_{-}  -\left(\vec{\phi}^*(\mathbf{r})\cdot\vec{d}\right)  a^{\dagger} \sigma_{-}\notag\\
&\equiv g a \sigma_{+} + g^{*} a^{\dagger}\sigma_{-}, 
\end{align}where $g\equiv -\vec{\phi}(\mathbf{r})\cdot\vec{d}$. We have omitted the latter two terms which give zero contribution for single-photon transport. For single-photon transport, the rotating wave approximation (RWA) becomes exact.

For a resonator that support a pair of degenerate WGMs, the electric field operator is given by
\begin{align}
\mathbf{E}(\mathbf{r}) &= a\vec{\phi}_a(\mathbf{r}) + a^{\dagger} \vec{\phi}_a^*(\mathbf{r}) + b\vec{\phi}_b(\mathbf{r}) + b^{\dagger} \vec{\phi}_b^*(\mathbf{r})\notag\\
&= \left(a+b^{\dagger}\right)\vec{\phi}(\mathbf{r}) + \left(a^{\dagger}+b\right)\vec{\phi}^{*}(\mathbf{r}),
\end{align}where we have used $\vec{\phi}_a(\mathbf{r})=\vec{\phi}_b^*(\mathbf{r})\equiv  \vec{\phi}(\mathbf{r})$. Thus, the interaction is given by
\begin{align}\label{E:GeneralInteraction}
H_I &= -\left[\left(a+b^{\dagger}\right)\vec{\phi}(\mathbf{r}) + \left(a^{\dagger}+b\right)\vec{\phi}^{*}(\mathbf{r})\right]\cdot\left(\vec{d}\sigma_{+} + \vec{d}^* \sigma_{-}\right)\notag\\
&= -\vec{\phi}\cdot\vec{d} \left(a+b^{\dagger}\right)\sigma_{+} -\vec{\phi}\cdot\vec{d}^* \left(a+b^{\dagger}\right)\sigma_{-}-\vec{\phi}^*\cdot\vec{d} \left(a^{\dagger}+b\right)\sigma_{+}-\vec{\phi}^{*}\cdot\vec{d}^{*} \left(a^{\dagger}+b\right)\sigma_{-}\notag\\
&=-\vec{\phi}\cdot\vec{d}\, a\sigma_{+} -\vec{\phi}\cdot\vec{d}^*\, b^{\dagger}\sigma_{-}-\vec{\phi}^*\cdot\vec{d}\, b\sigma_{+}-\vec{\phi}^{*}\cdot\vec{d}^{*}\, a^{\dagger}\sigma_{-}\notag\\
&\equiv \left(g_a a\sigma_{+} + g_a^* a^{\dagger}\sigma_{-}\right) + \left(g_b b\sigma_{+} + g_b^* b^{\dagger}\sigma_{-}\right),
\end{align}where $g_a \equiv -\vec{\phi}\cdot\vec{d}$, and $g_b \equiv -\vec{\phi}^*\cdot\vec{d}$. This gives the most general form of the interaction for a WGM-type resonator interacting with a two-level atom. Note that in systems without underlying time-reversal symmetry, $|g_a| \neq |g_b|$.

Here we make further remarks on the form of the interaction, Eq.~(\ref{E:GeneralInteraction}):
\begin{enumerate}
\item If the transition from state $|-\rangle$ to $|+\rangle$ corresponds to a $\Delta m=0$ of a real atom ($m$ is the magnetic quantum number), $\vec{d}$ is a real vector. On the other hand, for an atomic $\Delta m=\pm 1$ transition, such as might be induced by circularly polarized light, $\vec{d}$ is necessarily a complex vector~\cite{Cohen-Tannoudji:1977,Allen:1987,Mandel:1995}. When $\vec{d}$ is a real vector, one has $g_a = g_b^* \equiv g$, and Eq.~(\ref{E:GeneralInteraction}) becomes
\begin{equation}
H_I = \left(g a\sigma_{+} + g^* a^{\dagger}\sigma_{-}\right) + \left(g^* b\sigma_{+} + g b^{\dagger}\sigma_{-}\right).
\end{equation}

For the kind of experimental setup in Refs.~[\onlinecite{Aoki:2006,Dayan:2008}], if the WGMs are TE mode so the electric field is linearly polarized along the $z$-axis (please refer to Fig.~\ref{Fi:PhiDefinition} for axis orientation), such an electric field will induce only $\Delta m=0$ transitions between atomic states. This yields a real dipole moment $\vec{d}$ and thus one has $g_a = g_b^*$. 

\item Other simplifications of Eq.~(\ref{E:GeneralInteraction}) can be made. For example, using the fact that in the cylindrical coordinate system $(\rho, \phi, z)$ one has
\begin{equation}
\vec{\phi} = \left(E_\rho, -i E_\phi, E_z\right) e^{-i m \phi},
\end{equation}(see Appendix~\ref{A:SymmetryTransformations}, or Ref.~[\onlinecite{Snyder:1983}]), and suppose that the electric dipole of the quantum dot is transverse to the $\hat{\phi}$ direction, and the azimuthal origin lying at the location of the quantum dot, the interaction takes the form
\begin{equation}
H_I = \left(g a\sigma_{+} + g^* a^{\dagger}\sigma_{-}\right) + \left(g b\sigma_{+} + g^* b^{\dagger}\sigma_{-}\right),
\end{equation}where $g_a=g_b\equiv g$. Such a choice was made in Ref.~[\onlinecite{Srinivasan:2007}]
\end{enumerate}

%For each case, the single-photon transport properties could be computed exactly using the procedures outlined in Sec.~\ref{Sec:Solutions}.

%\section{Reflection and couter-propagating WGM excitation spectrum}\label{A:ReflectionAndWGM}

%Here we plot the reflection spectrum $R\equiv {|r(\omega)|}^2$ and the counter-propagating WGM excitation spectrum ${|e_b|}^2$ in Fig.~\ref{Fi:RArray}, and Fig.~\ref{Fi:EbArray}. The parameters are the same as those in the transmission spectrum plot in Fig.~\ref{Fi:TArray}. 

\section{Waveguide-ring resonator system}\label{A:List}

Before coupling to the atom, one has to calibrate the system of waveguide-resonator, and to quantify the physical parameters such as the coupling strength, the intrinsic loss, and the $Q$-factor of the resonator. Moreover, the waveguide-ring resonator system itself has been widely adapted in opto-electronic devices to achieves functions such as add-drop filter, delay line, and signal switching.  In this appendix we give the transmission amplitude for this important special case.

The transmission amplitude of a system of waveguide-ring resonator is given by
\begin{equation}\label{E:WRt}
t(\omega)=\frac{\left(\omega-\omega_c+i\frac{1}{\tau_c}\right)^2-{|h|}^2+\Gamma^2}{\left(\omega-\omega_c+i\frac{1}{\tau_c}+i\Gamma\right)^2-{|h|}^2},
\end{equation}which can be easily obtained by either starting with the Hamiltonian $H$ in Eq.~(\ref{E:Hamiltonian}) and the state $|\epsilon^{+}\rangle$ in Eq.~(\ref{E:InteractingEigenstate}) by omitting the atomic parts; or from the expression of $t$ in Eq.~(\ref{E:th}) by letting 
the atomic transition frequency $\Omega\rightarrow\infty$ and 
the resonator-atom coupling $g\rightarrow 0$, which effectively decouples the atom from the waveguide-resonator part.  Note only the magnitude of $h$ appears in the amplitude, its phase does not. For simplicity, we assumed $V_a = V_b \equiv V$, and $g_a = g_b =g$. 

For the special case when $h=0$, Eq.~(\ref{E:WRt}) and other amplitudes yield the well-known simple forms:
\begin{subequations}\label{E:WRAmpNoh}
\begin{align}
t(\omega) &=\frac{\omega-\omega_c+i\frac{1}{\tau_c}-i\Gamma}{\omega-\omega_c+i\frac{1}{\tau_c}+i\Gamma},\label{E:WRtNoh}\\
e_a(\omega)&=\frac{V^*}{\omega-\omega_c+i\frac{1}{\tau_c}+i\Gamma},\\
r(\omega)&=0=e_b(\omega).
\end{align}
\end{subequations}Note that the reflection amplitude $r$ is always zero even when the transmission amplitude is not of magnitude 1. Only the phase-matched resonator WGM ($a$) is excited. It is well-known that the ring resonator acts as an all-pass filter when the resonator's intrinsic loss ($1/\tau_c$) is negligible such that $1/\tau_c \ll \Gamma$. In this limit, $|t|\simeq1$ for all frequency $\omega$. On the other hand, when the resonator intrinsic loss dominates such that
 %$1/\tau_c \gg \frac{{|V|}^2}{2 v_g}$,  
 $1/\tau_c \gg \Gamma$, one again has $|t|\simeq1$ for all frequency $\omega$, and the ring resonator again acts as an all-pass filter. %When the intrinsic loss of the resonator is tunable, the This is analogous to Q-switching.

On the other hand, for the lossless case $1/\tau_c=0$, the locations of the local minima and maxima of the transmission spectrum can be obtained by solving for $\omega$ from $\frac{d{|t|}^2}{d\omega}=0$, which yields
\begin{equation}
\omega = \omega_c, \quad \omega_c \pm \sqrt{{|h|}^2-\Gamma^2}\equiv \omega_{\pm}.
\end{equation}
The number of dips in transmission thus depends upon the sign of 
\begin{equation}\label{E:hCoupling}
{|h|}^2 - \Gamma^2,
\end{equation}\emph{i.e.}, if $|h|>\Gamma$ the spectrum exhibits a doublet structure; while if $|h|<\Gamma$, the spectrum shows a single dip.

\section{Reflection and couter-propagating WGM excitation spectrum}\label{A:ReflectionAndWGM}

\begin{figure}[htb]
%\scalebox{1.4}{\includegraphics{R_Array.eps}}
\scalebox{1}{\includegraphics[width=\columnwidth]{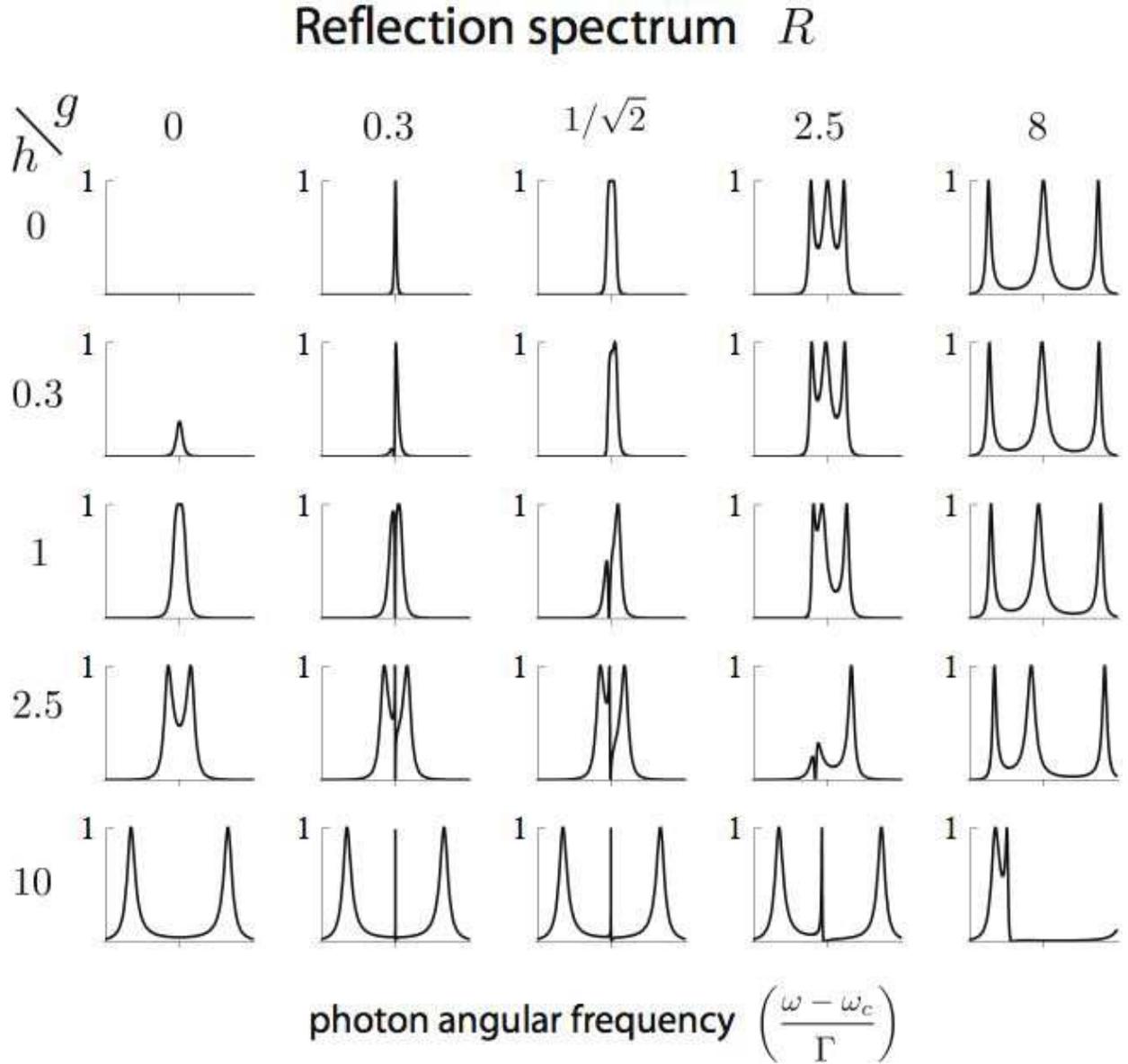}}
\caption{Map of the reflection spectrum $R$ as $g$ and $h$ are varied. $h$ is real for this map. Also, $g_a = g_b$ and $g\equiv |g_a|$. $R+T=1$. The location of $\omega=\omega_c$ is indicated by the gray tick in each figure.}\label{Fi:RArray}
\end{figure}

\begin{figure}[htb]
%\scalebox{1.4}{\includegraphics{Eb_Array.eps}}
\scalebox{1}{\includegraphics[width=\columnwidth]{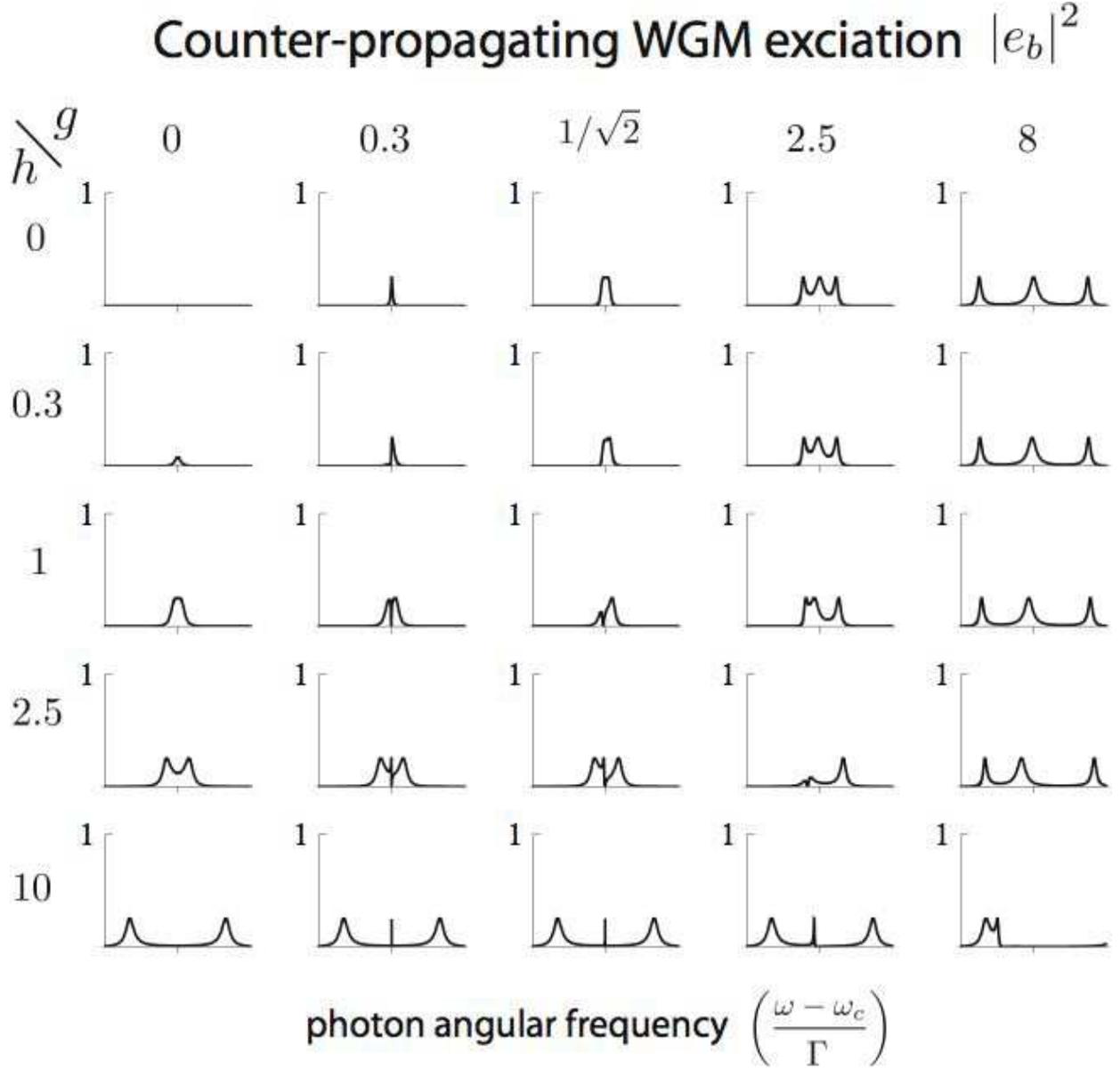}}
\caption{Map of the counter-propagating WGM excitation ${|e_b|}^2$ as $g$ and $h$ are varied. $h$ is real for this map. Also, $g_a = g_b$ and $g\equiv |g_a|$. The normalized spectrum $\frac{\Gamma}{2 v_g} {|e_b|}^2$ is plotted. The location of $\omega=\omega_c$ is indicated by the gray tick in each figure.}\label{Fi:EbArray}
\end{figure}

\begin{figure}[htb]
\scalebox{1}{\includegraphics{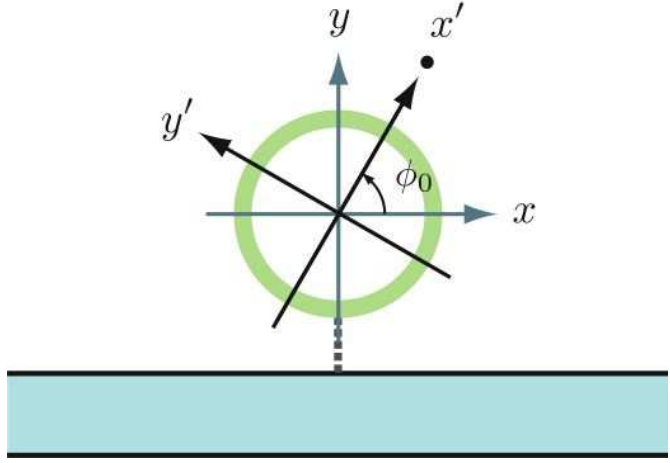}}
%\scalebox{1}{\includegraphics[width=\columnwidth]{PhiDefinition.eps}}
\caption{(Color online) The orientation between the $x'$-$y'$ and $x$-$y$ coordinate systems. The angle between the $x'$-axis and $x$-axis is $\phi_0$. The angle between the $x'$-axis and the mirror-plane ($y$-$z$ plane) is thus $\bar{\phi}\equiv \pi/2-\phi_0$.}\label{Fi:PhiDefinition}
\end{figure}

Here we plot the reflection spectrum $R\equiv {|r(\omega)|}^2$ and the counter-propagating WGM excitation spectrum ${|e_b|}^2$ in Fig.~\ref{Fi:RArray}, and Fig.~\ref{Fi:EbArray}. The parameters are the same as those in the transmission spectrum plot in Fig.~\ref{Fi:TArray}.

\section{Waveguide-Ring Resonator-Atom System at $h=0$}\label{A:hZero}

%The inter-mode backscattering $h$ originates from the fabrication imperfections, and is normally uncontrollable and unwanted. There has been tremendous experimental efforts to minimize the inter-mode backscattering. 
In this section, we give the results wherein the inter-mode backscattering is not present. The results are written in the form revealing the underlying physics.

When $h=0$, solving Eqs.~(\ref{E:EoM21})-(\ref{E:EoM25}) for $Q$, $t$, $r$, $e_a$, $e_b$, and $e_q$ gives:
\begin{subequations}
\begin{align}\label{E:tk}
Q &= \frac{\omega-\omega_0}{v_g},\\
t & = \frac{\left(\omega - \Omega + i \frac{1}{\tau_q}\right)\left(\omega-\omega_c + i \frac{1}{\tau_c}\right)-2|g|^2}{\left(\omega - \Omega + i \frac{1}{\tau_q}\right)\left(\omega-\omega_c + i \frac{1}{\tau_c}+ i \Gamma\right)-2|g|^2}-\frac{i \Gamma}{\omega-\omega_c+ i \frac{1}{\tau_c}+ i \Gamma},\label{E:t}\\
r &=e^{2i \theta_g}\left[\frac{- i (\omega-\Omega+ i\frac{1}{\tau_q}) \Gamma}{(\omega - \Omega + i \frac{1}{\tau_q})(\omega-\omega_c + i \frac{1}{\tau_c}+ i \Gamma)-2|g|^2}+\frac{i \Gamma}{\omega-\omega_c+ i \frac{1}{\tau_c}+ i \Gamma}\right],\label{E:r}\\
e_a &=\frac{(\omega-\Omega+i\frac{1}{\tau_q}) \frac{V^*}{2}}{(\omega - \Omega + i \frac{1}{\tau_q})(\omega-\omega_c + i \frac{1}{\tau_c}+ i \Gamma)-2|g|^2}+\frac{\frac{V^*}{2}}{\omega-\omega_c+ i \frac{1}{\tau_c}+ i \Gamma}\label{E:ea},\\
e_b &=e^{2 i \theta_g}\left[\frac{(\omega-\Omega+i\frac{1}{\tau_q}) \frac{V^*}{2}}{(\omega - \Omega + i \frac{1}{\tau_q})(\omega-\omega_c + i \frac{1}{\tau_c}+ i\Gamma)-2|g|^2}-\frac{\frac{V^*}{2}}{\omega-\omega_c+ i \frac{1}{\tau_c}+ i \Gamma}\right]\label{E:eb},\\
e_q &=\frac{gV^*}{(\omega - \Omega + i \frac{1}{\tau_q})(\omega-\omega_c + i \frac{1}{\tau_c}+ i \Gamma)-2|g|^2}\label{E:eq},
\end{align}
\end{subequations}
which are valid in both strong and weak coupling regimes. $g\equiv |g| e^{i \theta_g}$. Note that although $t$, $e_a$, and $e_q$ depend upon the magnitude of $g$ only, both $r$ and $e_b$, however, are proportional to $g^2$ and thus depend upon the phase of $g$ as $\sim e^{2 i \theta_g}$. Only when $h=0$, can one eliminate the phase dependence on $g$ by properly choosing the azimuthal origin of coordinate system.  The above expressions give explicitly the phase dependence without making the assumption that $g$ is real.

The form of the amplitudes, Eqs.~(\ref{E:t})-(\ref{E:eq}), is a direct consequence of the fact that the ring resonator supports two degenerate whispering-gallery modes. Two orthogonal non-propagating modes can be formed as linear superpositions of these two degenerate modes: one of them has non-zero amplitude at the location of the atom, and gives a contribution which resembles that of the waveguide-single-mode cavity-atom system~\cite{Unpublished}, but with an effective atom-resonator coupling $\sqrt{2} g$ and an effective resonator-waveguide coupling $|V|/\sqrt{2}$. The other mode on the other hand has zero amplitude at the location of the atom, and thus decouples from the atom. This mode gives a contribution which resembles that of the waveguide-single-mode cavity system, with an effective resonator-waveguide coupling $|V|/\sqrt{2}$. This explains the form of the amplitudes.

\pagebreak
\newpage

\pagebreak
\newpage

\pagebreak
\newpage

\pagebreak
\newpage

\pagebreak
\newpage

\pagebreak
\newpage

\pagebreak
\newpage

\pagebreak
\newpage

\pagebreak
\newpage

\pagebreak
\newpage

\pagebreak
\newpage

\pagebreak
\newpage

\pagebreak
\newpage

\pagebreak
\newpage

\pagebreak
\newpage

\pagebreak
\newpage

\end{document}